\documentclass[manuscript]{aastex}
\newcommand{\myemail}{leon@kurp.hut.fi}
\usepackage{longtable}
\usepackage{color}
\usepackage{enumerate}
\usepackage{natbib}
\usepackage{amsmath}
\usepackage{lscape}
\slugcomment{ Accepted to ApJ: April 14, 2012}

\shorttitle{The (sub-)mm -- $\gamma$-ray   connection.}
\shortauthors{Le\'on-Tavares et al.}

\def\fermilat{\textit{Fermi}/LAT}
\def\fermi{\textit{Fermi}}
\def\planck{\textit{Planck}}

\begin{document}
\title{Exploring the relation between (sub-)millimeter radiation  and $\gamma$-ray emission in blazars  with  \emph{Planck} and \emph{Fermi}} 

\author{J. Le\'on-Tavares\altaffilmark{1}, E.  Valtaoja\altaffilmark{2},  P. Giommi\altaffilmark{3},  G. Polenta\altaffilmark{3,4}, M. Tornikoski\altaffilmark{1}, A. L\"ahteenm\"aki\altaffilmark{1}, D. Gasparrini\altaffilmark{3}, S. Cutini\altaffilmark{3} }

\altaffiltext{1}{Aalto University Mets\"ahovi Radio Observatory,  Mets\"ahovintie 114, FIN-02540 Kylm\"al\"a, Finland; \myemail{}}
\altaffiltext{2}{Tuorla Observatory, Department  of Physics and Astronomy, University of Turku, 20100 Turku, Finland}
\altaffiltext{3}{ASI Science Data Center, ASDC c/o ESRIN, via G. Galilei, 00044 Frascati, Italy}
\altaffiltext{4}{INAF - Osservatorio Astronomico di Roma, via di Frascati 33, 00040, Monte Porzio Catone, Italy}

\begin{abstract} 
The coexistence of  \planck\ and \fermi\  satellites in orbit has enabled the exploration of  the  connection between the  (sub-)millimeter and    $\gamma$-ray emission  in a  large sample of  blazars. We find  that  the $\gamma$-ray emission and the (sub-)mm luminosities are correlated over five orders of magnitude, $L_{\gamma}~\propto~L_{(sub-)mm}$. However, this correlation is not significant at some frequency bands when simultaneous observations are considered. The most significant statistical correlations, on the other hand, arise  when observations are quasi-simultaneous  within 2 months.  Moreover, we find that sources  with   an approximate spectral turnover in the middle of the mm-wave regime are more likely to be strong  $\gamma$-ray emitters.  These results suggest a physical relation between the newly injected  plasma components in the jet and the high levels of  $\gamma$-ray emission. 
  \end{abstract}

\section{Introduction}

Within the three years of its operation,  the \emph{Fermi} satellite  has confirmed that  the extragalactic  $\gamma$-ray sky is  dominated by emission  from blazars \citep{2fgl,2lac}. A blazar is an unusual type of active galactic nuclei  (AGN) in which a relativistic jet points toward Earth, and due to relativistic effects the emission is amplified and variability time scales  look shorter. Blazars come in two flavours, with prominent  or weak   broad-emission  lines, and are called Flat Spectrum Radio Quasars (FSRQ) or  BL~Lac objects (BLLacs), respectively. 

 However, during the \fermi\ era it has been shown that not all blazars are strong $\gamma$-ray emitters \citep[e.g.][]{lister_2011, pep_asdc}, and the likelihood of detection is probably related to faster  and  brighter jets \citep[e.g.][]{kovalev_2009, nieppola_2011},  jets pointing closer to our line of sight \citep{lister_2009},  larger  apparent jet opening angles  \citep{pushkarev_2009,ojha_2010}, higher variability  \citep[e.g.][]{richards_2011}, or the presence of multiple inverse-Compton components \citep{fermi_sed_2010,turler_2011}.  In addition, it has been shown that the highest levels of  $\gamma$-ray emission are closely related to  ejections of superluminal components \citep[e.g.][]{agudo_2011} and ongoing  high-frequency radio  flares \citep{leontavares_2011_flares}.   Whether a blazar needs to fulfill  all (or a specific combination) of the above conditions in order to  radiate in $\gamma$-rays  is  one of the most important  questions that  still needs to be answered about the nature of $\gamma$-ray emission in blazars.

The  complete diagnostic of what makes a blazar bright at $\gamma$-ray wavelengths   requires  measurements of the so far poorly  explored  (sub-)mm spectral bands. Because mm and sub-mm radiation in blazars merely samples the optically thin regime of  the synchrotron spectrum,  thermal emission (from accretion processes) and radiation from lobes  is negligible in these bands \citep{giommi_2009}. Therefore,  (sub-)mm  measurements can efficiently probe the inner  regions of   blazar jets and shed light on the region where the bulk of the $\gamma$-ray emission is produced.

 In this work we compare  \fermi/LAT   $\gamma$-ray  photon fluxes   integrated over three different periods of time with  \emph{Planck} (sub-)mm  observations of   a  sample of   blazars.   The data are presented in Section 2,  and in Section 3 we perform a correlation analysis between intrinsic luminosities in both energy regimes.  Section 4  presents an analysis of mm/sub-mm spectral shapes  and $\gamma$-ray brightness. Our results are discussed and summarized in Section 5. Throughout this manuscript, we use a $\Lambda$CDM cosmology with values within $1\sigma$ of the WMAP results \citep{komatsu_2009}; in particular,   H$_{0}$=71 km s$^{-1}$ Mpc$^{-1}$, $\Omega_{m}=0.27$, and $\Omega_{\Lambda}=0.73$.

\section{The data}
We build our analysis on the sample of 105 blazars presented in \citet{pep_asdc}, who considered three samples of blazars with flux limits in  the soft X-rays (count-rates$_{\ 0.1-2.4\hspace{0.05cm}keV}>$ 0.3 counts s$^{-1}$), hard X-rays (S$_{\ 15-150\hspace{0.05cm}keV} > 10^{-11}$ erg cm$^{-2}$ s$^{-1}$), and $\gamma$-ray \footnote{ selected from the Fermi/LAT  Bright source list \citep{FBSL} with test statistics (TS) $> 100$ and Galactic latitudes larger than 10 degrees. } bands with additional 5 GHz flux density limits\footnote{ soft X-rays, S$_{5 GHz} > 200$ mJy;  hard X-rays,  S$_{5 GHz} > 100$ mJy; $\gamma$-rays, S$_{5 GHz} > 1$ Jy}. The advantage of using this sample is the availability of $\gamma$-ray flux   measurements (quasi-)simultaneous  to  \planck\ observations.  ESA's space mission \planck\ has been surveying the sky at nine frequencies  since August 2009 \citep{planck_mission}. Its payload includes two instruments: the Low Frequency Instrument (LFI) operating at   30, 44 and 70 GHz while the High Frequency Instrument (HFI) observes at 100, 143, 217, 353, 545   and  857 GHz.  The (sub-)mm flux densities  employed in this work are listed in Table 6 of \citet{pep_asdc}.

 The  \fermilat\ $\gamma$-ray photon fluxes   used in this study  are based on data in the  100~MeV to 300~GeV energy range   and the data processing procedure has been   fully described  in  section 3  of  \citet{pep_asdc}. For those sources where the Test Statistics (TS) in the whole energy band (0.1 - 300 GeV) delivers a value smaller than 25, an upper limit is given instead.   Our gamma -ray data   consist of three datasets  which we describe below: 
 
 \begin{itemize}
 
     \item  Simultaneous ($ <S_{\gamma} >_{sim}$):  The $\gamma$-ray photon flux was integrated over the period of time within which the source was observed at all \planck\  frequencies. A source in our sample typically sweeps over the \planck\ focal plane in about 2 weeks. Therefore,  $\gamma$-ray photon fluxes  integrated during the \planck\ observation can be considered  as   \emph{simultaneous} within one week.

 \item  Quasi-simultaneous ($ <S_{\gamma} >_{qua}$): The  $\gamma$-ray photon flux densities were obtained by integrating   \fermi\ data  over a period covering two months and centered on the Planck observing period, i.e. about one month before and  one month after the source was observed by \planck\ .  We consider these $\gamma$-ray data as  \emph{quasi-simultaneous} to the \planck\ observations on monthly timescales.

     \item  Average ($ <S_{\gamma} >_{ave}$): The $\gamma$-ray photon flux was integrated  over a long-term period (27 months, from August 2008 to November 2010). We  refer to these photon fluxes as \emph{average} $\gamma$-ray fluxes.
 
 \end{itemize}
   
  The number of  $\gamma$-ray detections  (N$_{det}$) in each of the above  Fermi datasets is as follows:  $ <N_{det} >_{sim}$ = 26,  $ <N_{det} >_{qua}$ = 56 and $ <N_{det} >_{ave}$ = 81.

 \begin{figure*}[!ht]
\includegraphics[width=\textwidth]{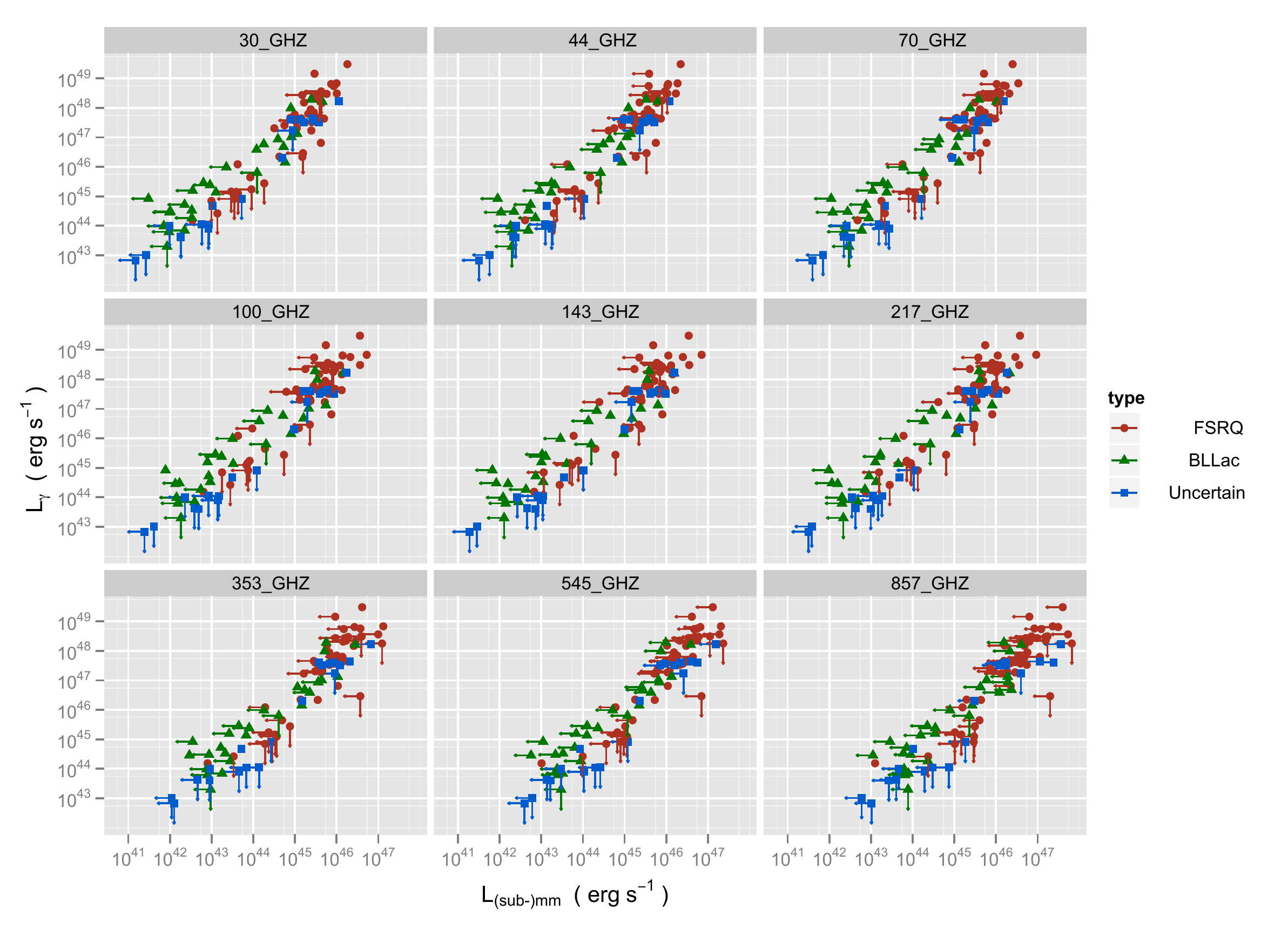}
\caption{\planck\ luminosities in (sub-)mm bands   versus  27 months average  \fermi\ luminosities.  Upper limits in the (sub-)mm and high-energy bands  are indicated by left and downward arrows, respectively. Red circles represent FSRQs, green triangles BLLacs, and blue squares sources with uncertain types. The partial correlation parameters  for each panel are given in Table \ref{tab:correlation}.}\label{fig_1}
\end{figure*}

  \begin{figure*}[!ht]
\includegraphics[width=\textwidth]{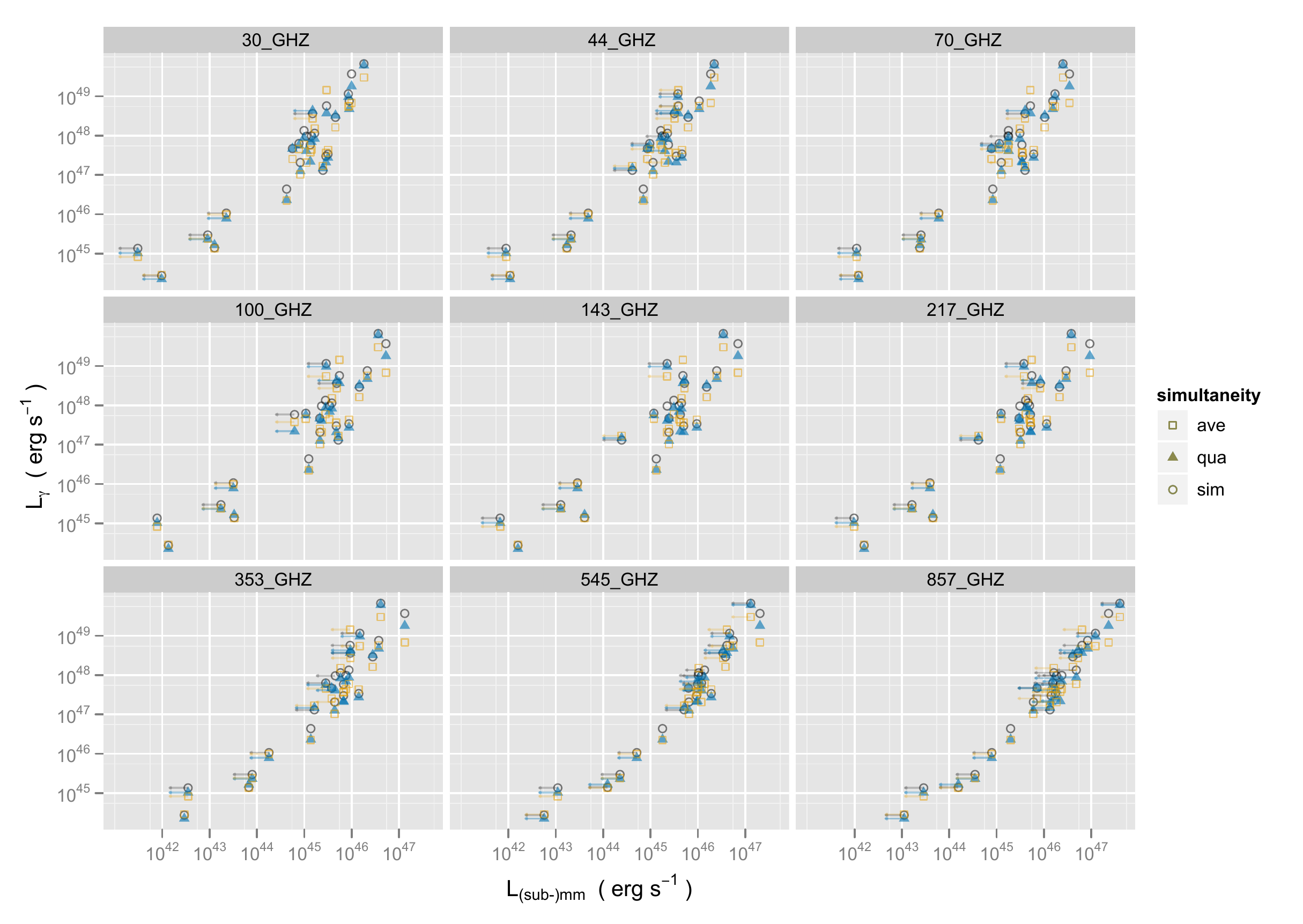}
\caption{\planck\ luminosities in (sub-)mm bands   versus   $\gamma$-ray luminosities using Fermi data averaged over three different periods of time: 2 weeks (sim), 2 months (qua) and 27 months (ave). The partial correlation parameters  for each panel are given in Table \ref{tab:sim}.}\label{fig_2}
\end{figure*}
 
\section{The (sub-)mm and $\gamma$-ray   emission  relationship}

In order to investigate the physical relation between the $\gamma$-ray and mm/sub-mm emission in blazars,  we look for  a statistical association between the wavelength regimes on the luminosity-luminosity plane.  The reason for performing the correlation analyses using  intrinsic luminosities  rather than  observed fluxes is that the flux-flux correlations will  be distorted  (and might even vanish) if  the coefficient $x$ in the relation $L_{\gamma} \propto L_{\nu} ^{x}$  is different from unity  \citep{feigelson_1983}.  On the other hand, if there is no  intrinsic luminosity-luminosity correlation then  no correlation will appear in the flux-flux plane as shown by \citet{feigelson_1983}. The mm/sub-mm luminosities have been computed according to the following expression,

\begin{equation}
L_{(sub-)mm} =   4 \pi d_{L}^{2} \frac{\nu S_{\nu}} {(1+z)} ~ \textrm{erg s}^{-1},
\end{equation}
where $d_L$ is the luminosity distance, $\nu$ corresponds to each of the nine  $Planck$ frequencies, and  $S_{\nu}$  is the single-epoch  flux density  taken from  \citet{pep_asdc}.

The $\gamma$-ray luminosity has been calculated  according to
\begin{equation}
L_{\gamma} =  4 \pi d_{L}^{2} \frac{S_{\gamma}(0.1,300)} {(1 + z) ^{1 -\alpha_{\gamma}}}  ~ \textrm{erg s}^{-1},
\end{equation}
where $\alpha_{\gamma} = \Gamma -1$,  $\Gamma$  being  the 27 months average photon spectral index and  $S_{\gamma} (0.1,300)$ is the photon energy flux  integrated between 0.1 and 300 GeV  calculated from the photon flux  ($F_{\gamma}$)  by  using the following expressions,
 \begin{equation}
S_{\gamma}(0.1,300)= 1.6 \times 10^{-4} ~\frac{\alpha_{\gamma} ~F_{\gamma}} {1-\alpha_{\gamma}} ~ [ 3000 ^{1-\alpha_{\gamma}} - 1] ~\hspace{0.5cm} \alpha_{\gamma} \neq 1  
\end{equation}
\begin{equation}
S_{\gamma}(0.1,300)=1.28 \times 10^{-3} ~ F_{\gamma} \hspace{0.5cm} \alpha_{\gamma}  = 1 
\end{equation}
derived from the relations presented in  \citet{ghisellini_2009}.

\subsection{Statistical methods}
 It is conceivable that the dependence on redshift  might induce artificial  correlations between the luminosities. For that reason  we must apply partial correlation methods to evaluate intrinsic $L_{\gamma} - L_{\nu}$ correlations.  Because of the presence of upper limits in   some of the $\gamma$-ray flux  and (sub-)mm flux density measurements in our sample,  survival analysis techniques are needed to properly quantify the correlation coefficient between these two emission bands.  Therefore  we have  applied  the Kendall's tau partial correlation for censored data \citep{akritas} to estimate whether there is an intrinsic correlation between luminosities once the influence of the  upper limits has been taken into account and the common dependence on redshift has been removed.

The FORTRAN code \texttt{cens\_tau}\footnote{http://www2.astro.psu.edu/statcodes/cens\_tau.f} implements the methods presented in \citet{akritas}   providing  a measure of the correlation between the two luminosities while excluding  the effects of the redshift ($\tau_{L L z}$). The probability that non-correlation exists between luminosities  (P$_{\tau}$) can be gleaned from the output of the code as well. We consider a  partial correlation significant if the probability of non-correlation is less than 5\% (P$_{\tau} < $ 0.05).

The slope of the correlations between (sub-)mm and $\gamma$-ray bands has been computed by the Akritas-Theil-Sen nonparametric regression  method implemented in the   routine  \texttt{cenken} of the R package \texttt{NADA}. This method is best suited to our purposes owing the presence of upper limits in both variables. 

\subsection{Significance test using the method of surrogate data }

To ensure that an intrinsic luminosity-luminosity correlation has not been drawn by chance we employed  surrogate data (uncorrelated  (sub-)mm and $\gamma$-rays luminosity pairs)  to quantify whether the correlation coefficient obtained by applying the censored partial correlation test can or cannot be reproduced by mere coincidence. Based on the surrogate method described in detail in \citet{surrogate}, we constructed  surrogate luminosity data sets by  following the next  procedure:

\begin{enumerate}
  
\item We start with a sample of $m$ sources with a firm measurement of their redshift. The (sub-)mm and $\gamma$-ray luminosities are thus estimated by using expressions  (1) and (2), respectively. Then, we run the censored partial correlation test to quantify the observed correlation strength ($\tau_{0}$).

 \item The sample is  split in $n$ redshift bins, each  bin should contain at least 10 sources. This criterion follows the arguments described in section 4 of \citet{surrogate}.

\item  In each bin we randomly shuffle the redshift and both  the (sub-)mm and  $\gamma-$ray  luminosities. The  censored status of the luminosities at each band -- either detected or upper limit -- sticks to the original luminosity value.

 \item  Using the surrogates from  each bin, we build an uncorrelated data set of $m$ pairs of luminosities with randomly assigned redshifts. Then we run the censored partial correlation test  to estimate the correlation coefficient $\tau_{i}$ and the probability that  the surrogate data set is non-correlated (P$_{i})$.
 
 \item We repeat the steps 3 and 4 a large number of times (N$_{trials}$)  and build the   distribution of  correlation coefficients for these surrogate uncorrelated luminosity pairs.

\end{enumerate}

 We use  the distribution of random correlation coefficients built on  N$_{trials} = 1000$ to compute the probability  of obtaining a significant correlation (P$_{i} < $ 0.05)  with a coefficient larger or equal (in terms of absolute value) than the  observed  one.  Then, the significance of the observed censored partial correlation (P$_{surrogate}$) is estimated by computing the ratio N$_{hits}/$N$_{trials}$, where N$_{hits}$ is  he number of times that we registered a significant correlation  with  $|\tau_{i}| \geq \tau_{0}$.
  
 An observed censored partial correlation  is considered real only if  the probability of getting  the same (or a larger) correlation coefficient is less than 5\% (P$_{surrogate} < 0.05 $). This criterion will ensure us  that observing a significant  (sub-)mm ad $\gamma$-ray luminosity-luminosity   partial correlation with a coefficient $\tau_{0}$ is not likely to occur by chance.

\subsection{The    L$_{\gamma}$ - L$_{(sub-)mm}$ correlation}

We  investigate the luminosity-luminosity correlation between the  (sub-)mm emission  and $\gamma$-ray fluxes in the sample  of blazars presented in \citet{pep_asdc}. We only consider sources with firm  measurements  of their redshift   and  $\gamma$-ray photon flux -- either detection or upper limit--  leaving us with a total of 98 sources. The luminosities at  the (sub-)mm and $\gamma$-ray regimes were computed according to expression 1 and 2, respectively.  In equation (2), $S_{\gamma}(0.1, 300)$ is the photon energy flux integrated over 27 months ($<S_{ave}>$). For those cases where non-detection of the sources was possible,   we estimate an upper limit of the luminosity  assuming  the average photon spectral  index for our sample, $<\Gamma >= 1.38$.  All luminosities are listed in Table \ref{tab:sample}.

Figure 1 shows the luminosity-luminosity relation between all (sub-)mm  frequency bands  and  average values of  $\gamma$-ray emission obtained by integrating Fermi data over 27 months.  In order to examine  the possible dependence  of the correlation on blazar type, the  various  blazar types  are symbol coded as shown in the  legend of  Figure 1.   FSRQs populate the upper right part of the luminosity-luminosity plane, whereas BLLacs and sources with uncertain type  continue the trend toward lower luminosities. It should  be noticed that some of the sources classified as ``uncertain type" are  actually known radio galaxies, however, for sake of consistency we keep the nomenclature as stated in Tables 1 to 3 of  \citet{pep_asdc}

The correlation parameters and their significances for each of the panels of Figure 1 are summarized in Table \ref{tab:correlation}.  The statistical methods described in the above subsections reveal that mm/sub-mm luminosities  are positively correlated with $\gamma-$ray luminosities over five orders  of magnitude when all source types are considered. However, differences arise when we compute partial correlations for FSRQs and BLLacs, separately.  The correlations between (sub-)mm  and $\gamma$-ray emission bands are significant and real when FSRQs are considered only. However, the surrogate test method does not provide evidence that correlations for BLLacs are significant.

The slope of the relation between (sub-)mm and  $\gamma$-ray luminosities  has been computed with the  Akritas-Theil-Sen nonparametric regression  method   and the fitted values  are also  listed in Table 1. For all 98 sources considered in our study,  the (sub-)mm - $\gamma$-ray luminosity relation can be  well approximated   as  $L_{\gamma} \sim L_{\nu} ^{x}$, where $<x_{All}>  = 1.27  \pm  0.02$. The slope  fitted for BLLacs  ($<x_{BLLac}>  = 1.13  \pm 0.03$) is somewhat shallower than the slope computed for FSRQ ($<x_{FSRQ}>  = 1.40  \pm  0.03$). While the different  photon spectral index  between FSRQs and BLLacs  \citep{ghisellini_2009} may play some role in the  difference between fitted slopes, the lack of correlation for BLLacs might indicate that different radiation mechanisms  are behind the $\gamma$-ray emission.     As we discuss in section 4, clean and well characterized samples of BLLacs are needed   to get further insight into the   physical conditions that make the difference between FSRQs and BLLacs.

The low detection rate at  the highest Planck frequencies can be a combination  of the high flux density detection  limit  at these frequencies -- see Table 3 in \citet{ercsc},  and the fact that blazars in general become brighter at submm  frequencies  only during flaring states \citep[e.g][]{marscher_1985,raiteri_2011}. Therefore, due to  the relatively low number of sources detected at 545 and 857~GHz the correlation  and best-fitted line parameters  for these frequency bands shown in Table \ref{tab:correlation} and \ref{tab:sim}   should be taken with caution.

 \subsection{Dependence of the L$_{\gamma}$ - L$_{(sub-)mm}$ correlation on simultaneity}

We  find that in our sample of blazars the $\gamma$-ray emission averaged over large periods of time (i.e. 27 months) is significantly correlated  to the (sub-)mm radiation.  This result  suggests a coupling of the emitting regions, however, to  get a physical insight about this relation  we next need to investigate whether the correlation  between (sub-)mm and $\gamma$-ray emission bands shows a trend for  strengthening  with simultaneity.  We emphasize again that our Fermi data set  contains information about the $\gamma$-ray emission averaged over three different periods of time, allowing us to  assess the intrinsic correlation between (sub-)mm  radiation and simultaneous, quasi-simultaneous and averaged $\gamma$-ray emission. 

To investigate  the dependence of  the $L_{\gamma}$ - $L_{(sub-)mm}$ correlation on simultaneity of the data, the levels of censoring at $\gamma$-rays must be controlled and therefore we only consider sources that were detected by Fermi at each of the three averaging periods.  Hereafter we refer to this subsample as the \emph{Fermi-detected} sample which comprises   24 sources. Most of them are classified as  FSRQ  type and based on their simultaneous SEDs all of them have been classified as low-synchrotron peaked  (LSP) in \citet{pep_asdc}.

Figure 2 shows the luminosity-luminosity relation for all (sub-)mm  frequency bands  and $\gamma$-ray emission integrated over three different periods of time. The simultaneity of the $\gamma$-ray emission to the Planck measurements  is symbol coded as shown in the legend at the right center of the plot.  Table \ref{tab:sim} reports  the  partial correlation parameters and associated  probabilities  for the three $\gamma$-ray data sets.

As can be seen from Table  \ref{tab:sim},  it is noticeable that the strongest  correlations   arise  when  \fermi\  measurements  have been integrated over long periods of time  (2 and 27 months), whereas the weakest and less significant correlations  are always obtained when simultaneous observations are involved. The remarkably stronger correlation  with the $\gamma$-ray photon fluxes integrated over 27 months is likely an artificial effect  due to averaging the $\gamma$-ray photon fluxes  over a very long period of  time. This reduces the  dynamical range of $\gamma$-ray photon flux observed  \citep{muecke_1997}, which in turn reduces the scatter of the $L_{\gamma}$-$L_{\nu}$  relation and improves the correlation strength.    Given the evidence that the strongest $\gamma$-ray flares tend to have a short duration time, with a typical timescale of a month \citep[see][]{Fermi_lc},  one might argue  that by averaging $\gamma$-ray data over two months we may be averaging  flares, which  would reduce the dynamical range  of   quasi-simultaneous $\gamma$-ray luminosities, and therefore stronger correlations may be induced. However, it is not likely that all sources considered in our study were flaring during the weeks  that  $Planck$  observed them. This leads us to believe that the  the most significant correlations  in our study  arise when using the  $\gamma$-ray photon fluxes averaged over 2 months.

\section{Shape of the synchrotron spectra and $\gamma$-ray brightness}

In this section we explore emerging trends between (sub-)mm spectral shape and  $\gamma$-ray brightness
with the aim to find out  whether the (sub-)mm spectrum shape can be considered as a new piece to the puzzle of what makes  a blazar shine at $\gamma$-rays. The radio to sub-mm spectra can be well approximated  by  two power laws \citep[e.g.][]{gear_1994, pep_metsahovi}, however the point where both power laws connect has always been assumed rather arbitrarily based on the spectral coverage of the  data sets involved.  This in turn might introduce some biases to the spectral indices fitted to parametrize the spectral shapes. 

Hence,  in this work we use a different approach  by fitting the Planck observations from 30 to 857 GHz with a broken power law model    where the  fiducial point where the intersection of the power laws occurs  is considered as a free parameter along the fit. The expression of the broken power law  used to model the (sub-)mm spectral range is given by,
  
\begin{equation}
 S(\nu) \propto \begin{cases} 
 \nu^{\alpha_{mm}} &\mbox{for} ~  \nu  \leq  \nu_{break} \\
 \nu_{break}^{(\alpha_{mm}-\alpha_{submm})} \nu^{\alpha_{submm}} & \mbox{for } \nu  > \nu_{break},
\end{cases} 
 \end{equation}
where $\alpha_{mm}$ and $\alpha_{submm}$ are the spectral indices for the mm and sub-mm part of the spectrum, respectively, and the two power laws are connected at the  break frequency $\nu_{break}$.

 To perform the analysis  of spectral shapes and $\gamma$-ray brightness, we select only those sources from \citet{pep_asdc} that have been detected in at least five Planck frequency bands, this with the aim  to produce a reliable modeling of the spectra. The 47 sources selected based on the criterion mentioned above, along with the best-fit model parameters, are listed in Table \ref{tab:fits}.

\begin{figure}
\includegraphics[width=\columnwidth]{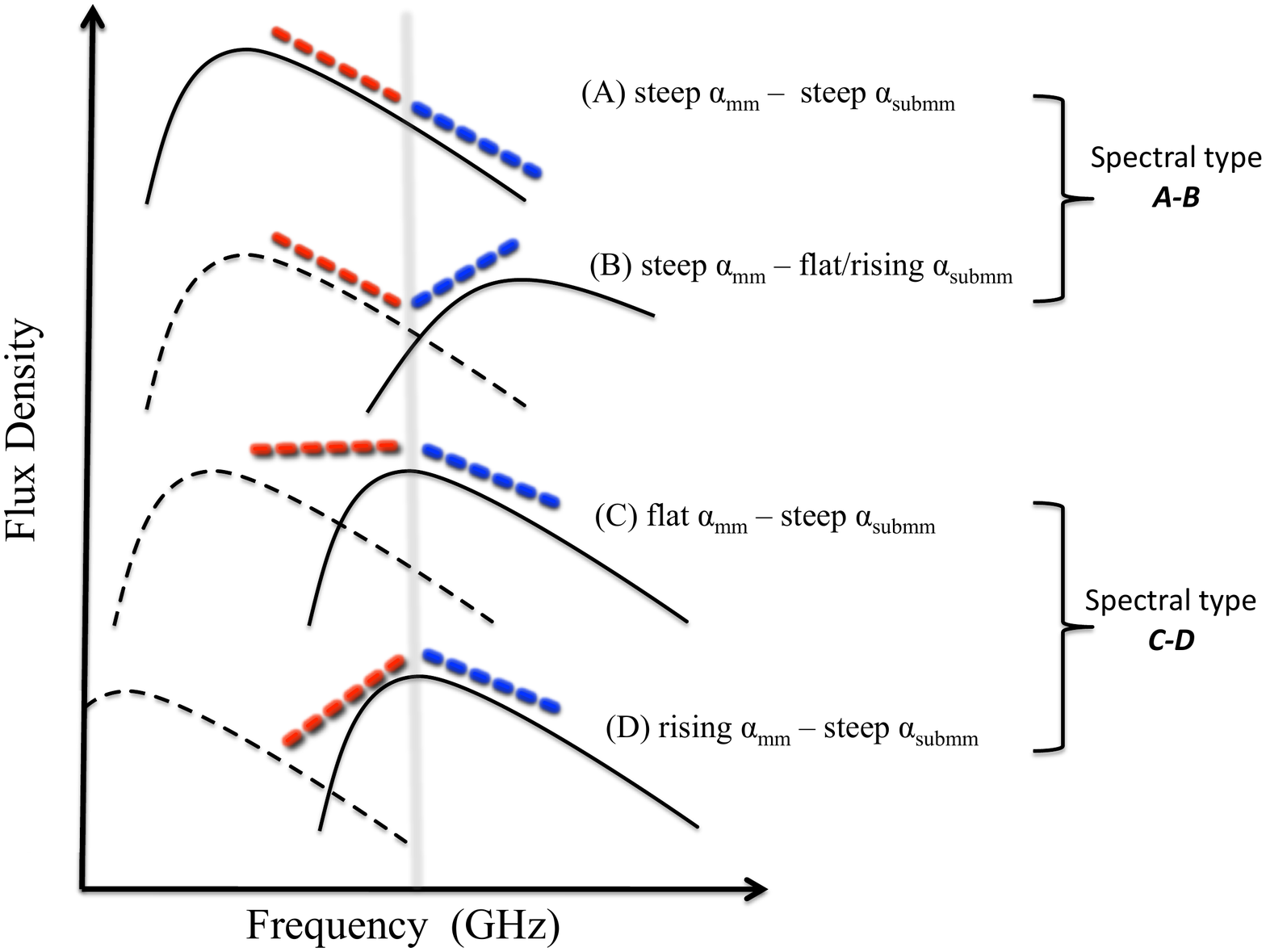} 
\caption{An illustration of all synchrotron spectral shapes we could possibly find in our sample, see text for detailed description. The overall spectrum has been approximated by   a broken power law with spectral indices $\alpha_{mm}$ (red) and $\alpha_{sub-mm}$ (blue), respectively. The  break frequency ($\nu_{break}$) is indicated by the grey thick vertical line. Aged synchrotron components are represented by dashed lines    }\label{fig:cartoon}
\end{figure}

The new (sub-)mm  observations provided by the Planck satellite  and  their modeling with a broken power law function allow us to typecast   the (sub-)mm spectral shapes in our sample.   Figure \ref{fig:cartoon}  shows  all four spectral shapes that could possibly be found in our sample. Next  we describe their characteristics and possible  interpretation in terms of  multicomponent synchrotron spectra:

\begin{enumerate}[(A)]

\item \emph{ Steep $\alpha_{mm}$ - Steep $\alpha_{submm}$:}  (sub-)mm emission  decreases monotonically with frequency. This straight spectral shape  from mm to sub-mm regimes is dominated by the underlying   emission   from  a  component  with a  synchrotron turnover frequency   located at cm wavelengths.  Quiescent activity levels  (i.e. no new young synchrotron components) is expected for  sources  featuring this spectral type.

\item \emph{Steep $\alpha_{mm}$ - Flat/rising $\alpha_{submm}$:}  mm spectrum is dominated by the emission from an   aged component that becomes self absorbed at the  cm regime, as in $(A)$. Despite that, a flat/rising $\alpha_{submm}$ spectrum betrays   the presence of a  new   component in an early development stage, located well within the radio core region. Then  a new flare connected to the ejection of a new VLBI component is likely to occur before the sub-mm spectrum steepens.  It is   possible  that a  rising $\alpha_{submm}$  could also be related to the presence of a dust component,  however,  a more detailed dissection of the spectra   is needed to  assess its presence.

\item \emph{Flat $\alpha_{mm}$ - Steep $\alpha_{submm}$:}  An overall  flatness of the  mm spectrum  can be explained by a succession of  several components in the jet  as stated in the so-called ``cosmic conspiracy" \citep{cotton_1980}. Nevertheless, the steepening of the  submm spectra  is consistent  with synchrotron  emission from a single component which  becomes self absorbed  somewhere in the  middle of the mm wavelength regime.

\item \emph{Rising $\alpha_{mm}$ - Steep $\alpha_{submm}$:} In the absence of adjacent strong spectral components
at lower or higher frequencies, a new component recently ejected from the radio core, with a self absorbed spectrum peaking in the mm regime, dominates the overall shape of the radio spectrum.  \end{enumerate}

The (sub-)mm spectra  of  the selected blazars were classified into the above spectral categories   based on the relative difference between  spectral indices   $\Delta\alpha = \frac{ \alpha_{submm} - \alpha_{mm}}  { \alpha_{mm} }$. The spectral classification for the 47 sources selected is reported  in  Table \ref{tab:fits} and  is defined by the following conditions:  

\begin{itemize}

\item   A : $\alpha_{mm} , \alpha_{submm}  < 0$ and $|\Delta\alpha|   < 0.5$;  relative difference between indices  is less than  50\%,

\item  B:    $\alpha_{mm} < 0$  and   $\alpha_{submm} \geq 0$,

\item   C :  $\alpha_{mm}, \alpha_{submm} < 0$ and $|\Delta\alpha|  > 0.5$;  relative difference between indices is greater than 50\%,

\item D:    $\alpha_{mm} >  0$ and $\alpha_{submm}  <  0$,

\end{itemize}

 The classification scheme described above  is sketched in Figure \ref{fig:cartoon}, which attempts to  compare the (sub-)mm spectral shapes of blazars  at different stages of the synchrotron components evolution. Based on successive approximations, we further group the four categories into  two  main spectral classes:  (i) \emph{\textbf{A-B}} includes  sources previously  classified  as A or B , these  represent the quiescent  and the very initial stages of an ejected component, respectively. (ii) The   spectral type \emph{\textbf{C-D}}  is conformed by   sources whose spectral shapes are categorized as type C or D, these specific spectral features  indicate the   presence of a  freshly ejected  synchrotron component where the spectral turnover is somewhere in the middle of the  mm wave regime.

\begin{figure}
\includegraphics[scale=0.6]{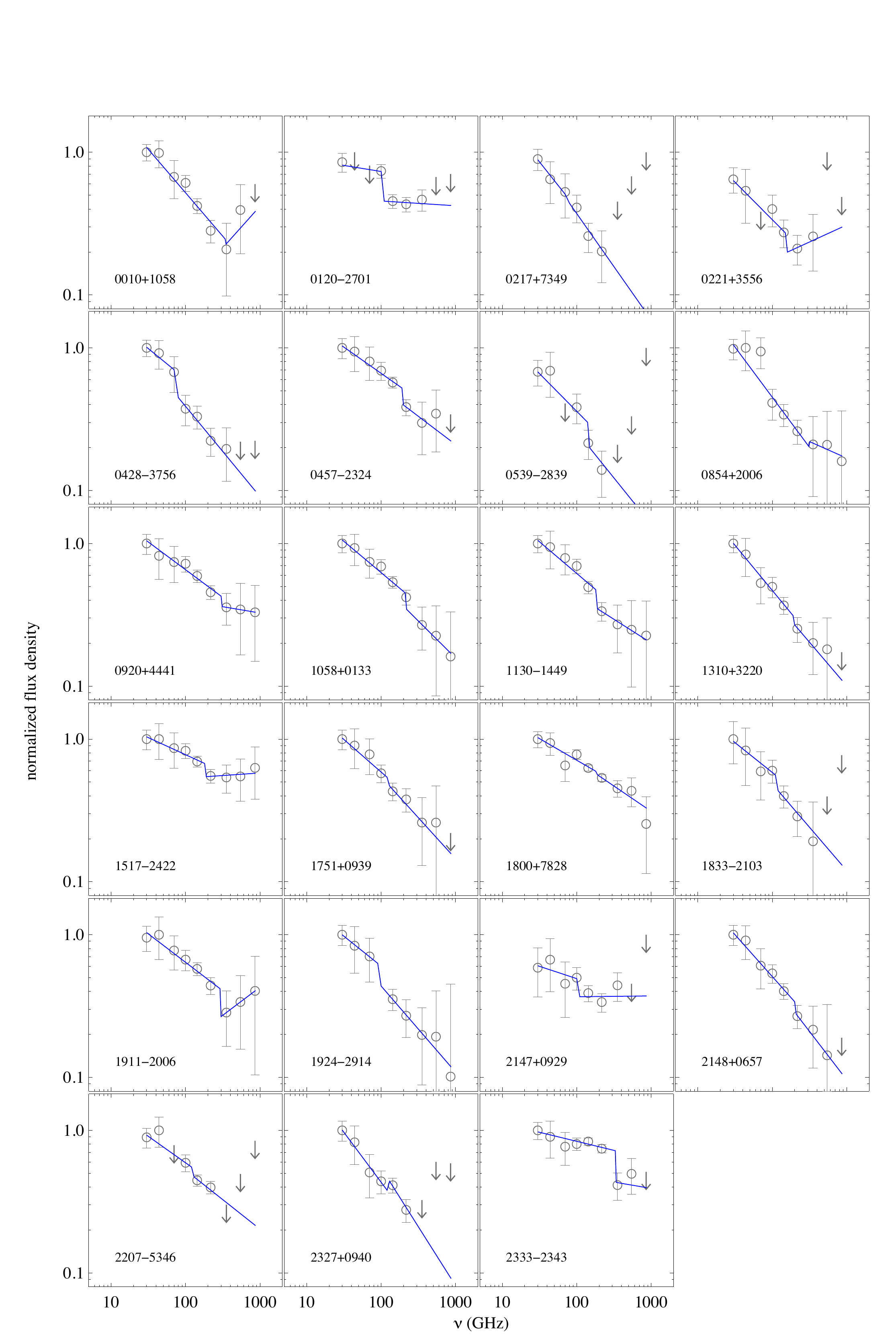} 
\caption{The (sub-)mm spectra for 23 sources classified within the spectral type \emph{A-B}. See Figure \ref{fig:cartoon} for a visual description of the types.   }\label{fig:sp_ab}
\end{figure}

\begin{figure}
\includegraphics[scale=0.6]{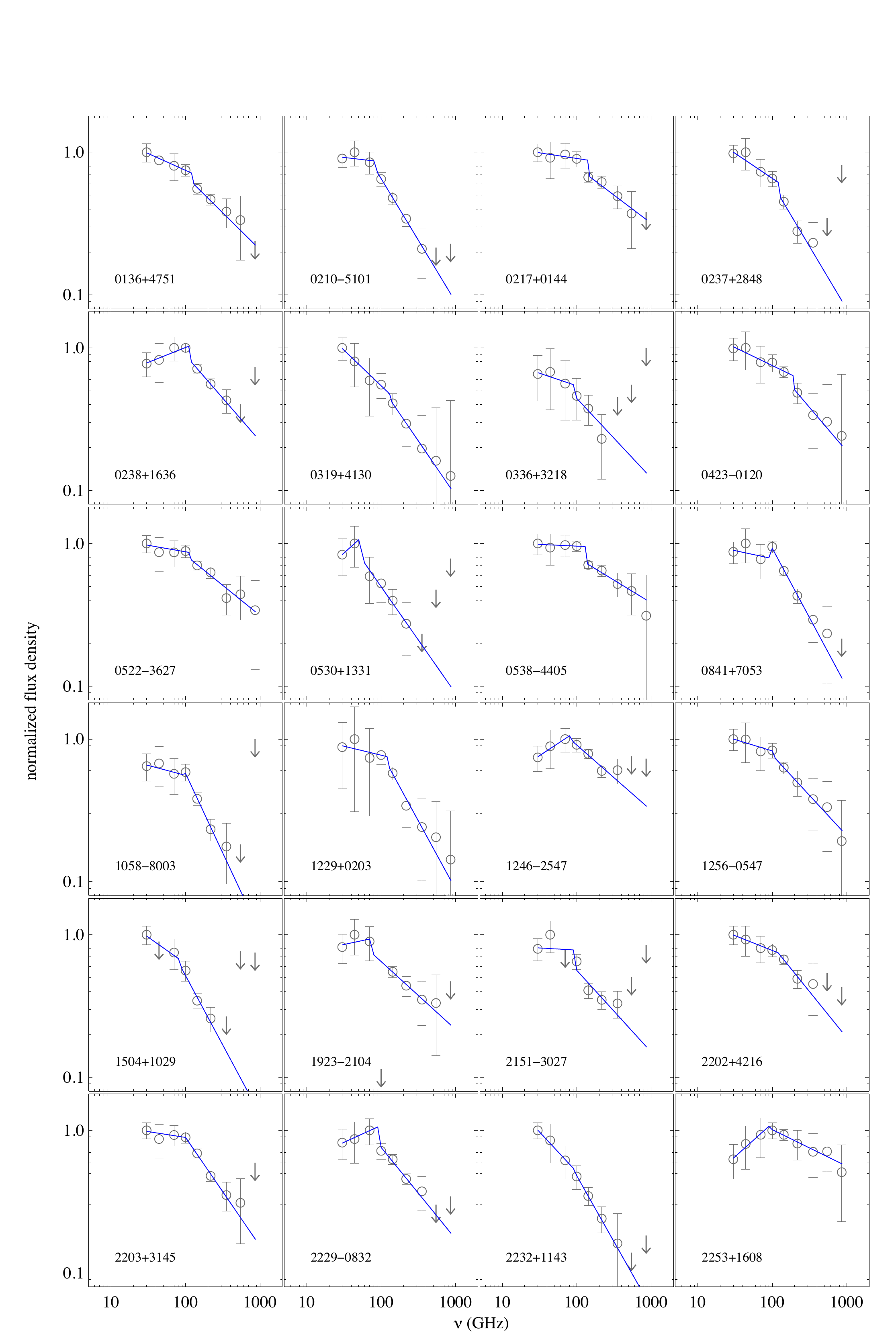} 
\caption{The (sub-)mm spectra for 24 sources classified within the spectral type \emph{C-D}. See Figure \ref{fig:cartoon} for a visual description of each spectral  type. }\label{fig:sp_cd}
\end{figure}

Based on the latter spectral type classification, we may classify our sample  into 23 sources of category  \emph{A-B} and 24 sources falling into the category of \emph{C-D}. Figures \ref{fig:sp_ab} and \ref{fig:sp_cd} show  the (sub-)mm spectra for both spectral types and overplotted are the best-fits to the (sub-)mm spectra using the broken power law model.

\begin{figure}
\includegraphics[width=\columnwidth]{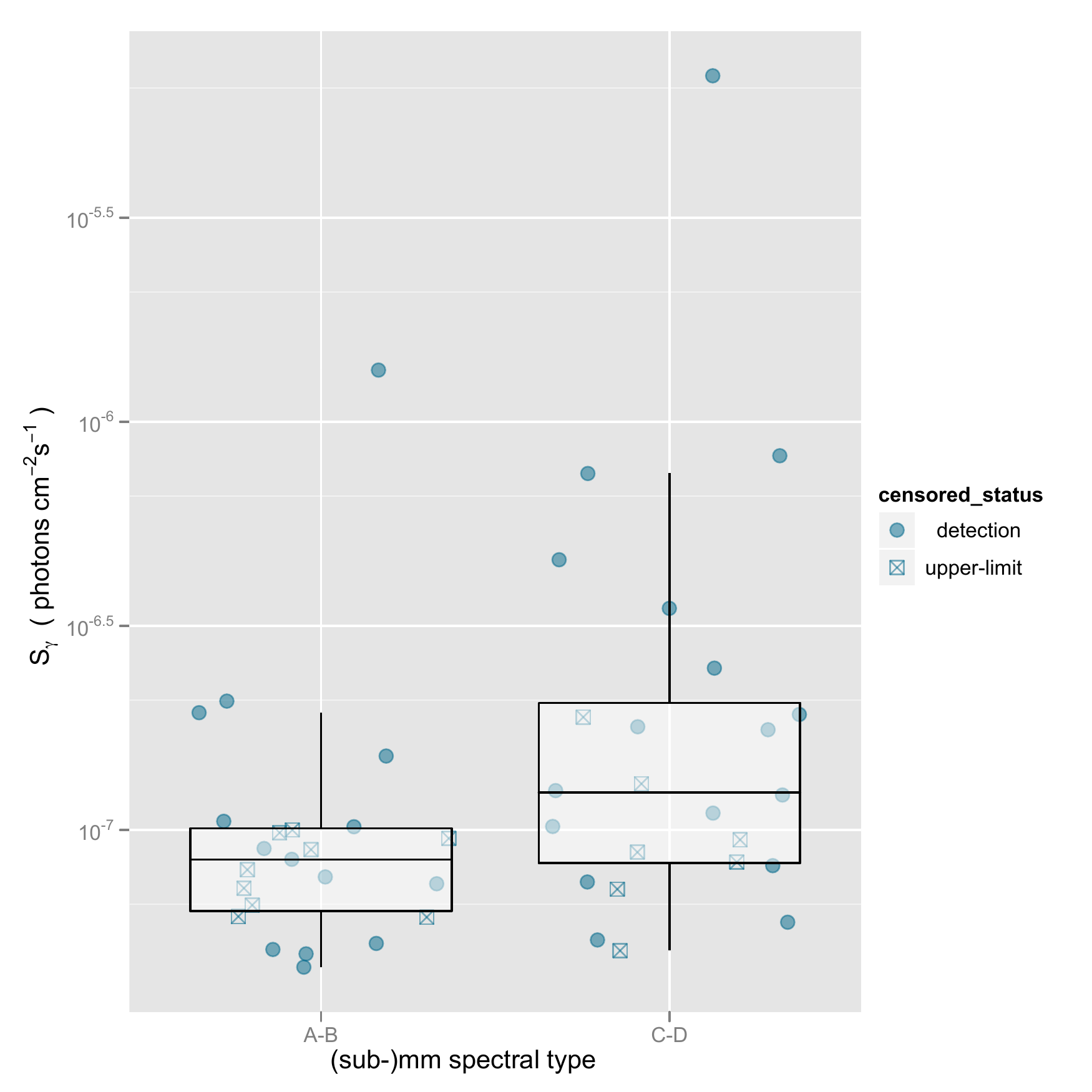} 
\caption{ The distribution of of $\gamma$-ray emission levels  for sources with spectral types \emph{A-B} (left) and \emph{C-D} (right)   are  shown in   boxplots, where the thick line at the middle of the box represents the median. The boxes comprise 75\% of the distributions. The levels of $\gamma$-ray emission for sources belonging to the type \emph{C-D} are significantly higher than for those classified as \emph{A-B}} \label{fig:boxplots}
\end{figure}

 Due to the fact that  $L_{\gamma} -L_{\nu}$ are best correlated when using quasi-simultaneous $\gamma$-ray fluxes ($<S_{\gamma}>_{qua}$), we proceeded to find out if there is a pattern in the spectral shapes that relates to the levels of quasi-simultaneous $\gamma$-ray emission.  Figure \ref{fig:boxplots}  shows the quasi-simultaneous $\gamma$-ray flux  distributions  as boxplots.  The levels of $\gamma$-ray emission for sources with (sub-)mm spectra belonging to the spectral classes A and B  (spectral type \emph{A-B}) are displayed in the  left boxplot of Figure \ref{fig:boxplots} while sources with spectra of the type C and D (spectral type \emph{C-D})  are shown in the boxplot to the right. Upper limits and detections are symbol coded as denoted in the legend of Figure \ref{fig:boxplots}.

  At first glance,  it seems that  sources  of the spectral type \emph{C-D}   are brighter at $\gamma$-rays than sources of the spectral type \emph{A-B}.  This can be gleaned from the size of the boxplots and the locations of the medians represented by the thick line  at the middle of the boxes displayed in Figure \ref{fig:boxplots}.  Differences in the distribution of  quasi-simultaneous $\gamma$-ray photon fluxes  for sources with  different (sub-)mm spectral types were statistically investigated. Because of  the presence of upper limits to the $\gamma$-ray photon fluxes we apply  the univariate two sample tests  implemented in the ASURV  package \citep{isobe_1986,asurv} to the $\gamma$-rays photon flux distributions displayed in Figure \ref{fig:boxplots}.   We further consider that  two populations  are significantly different if   the probability that the differences between populations  arise by  chance is   P $\leq 0.05$.   
  
 All five two-sample tests implemented in ASURV (Gehan's generalized Wilcoxon test -- permutation variance, Gehan's generalized Wilcoxon test  -- hypergeometric variance, logrank test, Peto-Peto generalized Wilcoxon test and Peto-Prentice generalized Wilcoxon test) converge to the same result:  the levels of  $\gamma$-ray emission of spectral types \emph{A-B} and \emph{C-D} are significantly different  and the probability that such difference arises by chance is less than 4\%. The statistical tests   show that the shapes  and medians of the distributions of $\gamma$-rays for spectral types \emph{A-B} and \emph{C-D}  are significantly different, where the levels of $\gamma$-ray emission for sources belonging to the type \emph{C-D} are significantly higher than for those classified as \emph{A-B}.  

We note that other classification schemes for blazars  have already been proposed based on the variability of the radio continuum spectra from cm to mm wavelengths \citep{angelakis_2011}. However, the classification scheme proposed in this work  -- based on successive approximations of the synchrotron components evolution -- for the fist time  permits to use the (sub-)mm spectral shapes as  flags for  high levels  of $\gamma$-ray emission.

\section{Discussion and summary}

While various studies during the Fermi-era  have found  evidence for a  correlation between radio and $\gamma$-rays  \citep[e.g.][]{kovalev_2009,surrogate,ghirlanda_2011,nieppola_2011}, our work  addresses for the first time this connection at  (sub-)mm wavelengths.  We have found a  significant correlation  between  (sub-)mm and $\gamma$-ray luminosities over five orders of magnitude. Since we have made use of partial-correlation, surrogate data methods and survival analysis techniques we are confident  that  the correlation  $L_{\gamma} \propto L_{\nu}$  is not an artifact of  the detection limits, chance  or  due to  the common dependence on redshift.  This result is consistent with previous studies \citep[e.g.][]{surrogate,ghirlanda_2011, nieppola_2011, arshakian_2012}, albeit using  high frequency radio observations. \citet{pep_asdc} do not report evidence of a significant  correlation between 143~GHz and $\gamma$-ray fluxes in  some of the sub-samples considered. This apparently contradiction with our findings is not  likely to be  influenced by  sources without redshift  not considered in this study. Instead, such difference  may be attributable to the fact that correlations  in the flux-flux plane can vanish if the slope of the luminosity-luminosity correlation is different from unity  (see discussion in Section 3) as  it turns out to be the case  for the  $L_{\gamma} \propto L_{(sub-)mm} $  relation (see Table \ref{tab:correlation}).

 The positive correlation between luminosities  is still robust   and  becomes even  stronger when only FSRQs are considered, however,  the correlation vanishes for BLLacs as a group.  In addition, the slope  of the fitted $L_{\gamma} \propto L_{\nu}^{x}$  relation is shallower for BLLacs.  The different gamma-radio behavior   between blazar types has been  found  in previous studies using 37~GHz data \citep{leontavares_2011_flares, nieppola_2011}.  On the other hand,  previous studies at low frequencies \citep{ghirlanda_2011, surrogate}  find that the radio-gamma correlation is significant for both FSRQs and BLLacs, separately. This apparent discrepancy might be related to the fact that  submm spectral indices  of FSRQs and BLLacs have been found to be significantly different   \citep{gear_1994}.  The latter result  was interpreted as an intrinsic difference in the underlying jets of BLLacs and FSRQ.  Unfortunately due to the low number of BLLacs   in the sample analyzed in section 4 --  see Table \ref{tab:fits}, we could not explore this  difference with the new (sub-)mm observation.

Nevertheless, before speculating into real physical differences between FSRQs and BLLacs, we would like to rise a  word of caution regarding the correlation results solely for the BLLac population. \citet{giommi_2012} have drawn attention to the fact that some sources classified as BLLacs might have been misclassified with FSRQ. This issue can be  a direct consequence of   the rather artificial criterion that draws the line between  FSRQs  and BLLacs  -- rest frame equivalent width (EW)   less than $ 5 \AA$ -- introduced twenty years ago \citep{stickel_1991}. It is plausible that a particular source was originally classified as BLLac  at the time when  its   non-thermal optical continuum  was prominent and high enough to swamp  any broad emission line in the optical spectral range,  as recently shown in \citet{padovani_2012}.  However,  after some time  the same source could have been classified as FSRQ when the optical continuum emission weakened, allowing the exhibition of broad emission lines.  To overcome this misclassification problem we should adopt a new  blazar taxonomical criterion to classify sources with intrinsically  prominent and weak  broad-line regions  as recently  proposed  in  \citet{ghisellini_2011} and \citet{ giommi_2012}.  Therefore  a compilation of a large sample  of  \textit{true} BLLacs  whose parent population includes  low-excitation radio galaxies (LERG) displaying FR-I morphologies, as proposed in \citet{giommi_2012},  is highly desirable  to draw robust  conclusions on the (sub-)mm and $\gamma$-ray connection in BLLacs.

We now turn to the dependence of the  $L_{\gamma} \propto L_{\nu}$ relation  on simultaneity.  Although the strongest correlation parameters are achieved when using $Fermi$ $\gamma$-ray photon fluxes averaged over 27 months, we  believe that such tight correlations are  a consequence of  averaging the $\gamma$-ray photon fluxes  over a very long period  of time. This reduces the scatter of the $L_{\gamma}$-$L_{\nu}$ relation   while  the detection of faint $\gamma$-ray sources    increases the $\gamma$-ray luminosity dynamical range, hence the correlation strengths are artificially enhanced. Furthermore, the $L_{\gamma} \propto L_{\nu}$  relation becomes  weak  if $\gamma$-ray and (sub-)mm data are obtained simultaneously. On the other hand, including quasi-simultaneous observations within monthly timescales  produces a  more significant correlation.   This might suggest an intrinsic delay between the emission in both wavebands, however, with the data currently available we cannot disentangle whether the (sub-)mm emission leads the production of $\gamma$-rays or vice versa. There is, however, recent evidence of  delays on monthly scales  between mm and $\gamma$-ray emission bands, the onset of a mm-flare occurring about 2 months before the maximum of the $\gamma$-ray emission is reached \citep{leontavares_2011_flares}.

Perhaps the most remarkable result  we find is that  brighter $\gamma$-ray sources tend to show a specific  shape in their radio spectra, i.e. flat/rising mm  and falling sub-mm spectral indices (spectral type \emph{C-D}), see Figure \ref{fig:cartoon} for the classification scheme proposed in this work.  This spectral shape is consistent with  a single synchrotron component becoming self absorbed in the  middle of the millimeter wavelength regime  that must be   responsible for  the last spectral turnover. 
Despite the fact that we have not attempted to properly  dissect the (sub-)mm spectra into single synchrotron components, turnover frequencies for  sources of the spectral type \emph{C-D} can be roughly constrained  to $\nu_{m} > $ 30 GHz.

Our finding that $\gamma$-ray emission is enhanced in sources of the spectral type \emph{C-D} -- featuring  an approximate spectral turnover at the middle of the mm wavelength regime -- suggests an important role of the plasma jet disturbances, responsible for shaping the (sub-)mm spectra,  in producing the   $\gamma$-ray emission.  Since  $B \propto \nu_{m}^{5} $  \citep{Lobanov_1998} sources of the spectral type \emph{C-D} ($\nu_{m} > 30$ GHz)  are expected to harbor strong magnetic fields in their cores leading to polarized emission.  This  argument  finds support in recent observational results by \citet{linford_2012}, where it was found that strong core polarization is related to  intense $\gamma$-ray emission.

The existence of large magnetic fields may reflect in either large electron densities in the emitting region  (in this case spectra must be dominated by emission from the core) or   the presence of   growing shocks emerging from the radio core region. To disentangle the nature  of the emission shown in the snapshot (sub-)mm spectra of Figures \ref{fig:sp_ab} and \ref{fig:sp_cd}, multifrequency VLBI monitoring is  highly desirable   to properly dissect the spectra into  individual components  \citep[e.g.] []{Lobanov_1999,Savolainen_2008} and to identify the contribution of the core and the emerging shocks to the overall radio spectrum. In addition,  upcoming Planck data releases will allow us  to build multi-epoch (sub-)mm spectra. Then, by  studying the evolution of the spectral shapes, we can get an insight into the changes of the physical properties of the jet (e.g., density, magnetic field, Doppler factor), and those must serve as input to theoretical models aiming to reproduce the levels of $\gamma$-ray emission observed.

In summary, we have used a joint sample of  soft X-ray, hard X-ray  and $\gamma$-ray selected blazars  to explore for the first time the relationship between (sub-)mm emission from 30 to 857 GHz and   $\gamma$-ray photon fluxes    integrated over three different periods of time: when a source was in the  Planck  field of view,  one month before and after the source was observed by Planck,   and  over 27 months.  Our findings can be listed as follows:

\begin{itemize}
   
\item  We have found a significant correlation between $\gamma$-ray and  (sub-)mm luminosities which holds over five orders of magnitude. Moreover, using a sample of bright $\gamma$-ray sources  detected by Fermi on each of the averaging periods mentioned above,  we  find  -- based on our statistical analysis  and bearing in mind that $\gamma$-ray photon fluxes averaged over long periods of time (i.e. 27 months)   enhance artificially the correlation strength --  that  quasi-simultaneous observations (within 2 months) showed the most significant statistical correlations.

\item Sources with high levels of $\gamma$-ray emission  show a characteristic signature in their radio spectra: flat/rising  mm  and steep sub-mm spectral indices (spectral type \emph{C-D} in Figure \ref{fig:cartoon}). This spectral shape can be associated with a single synchrotron component which becomes self absorbed in the middle of the mm wavelength regime. Such high spectral turnover frequencies  reveal the presence of emerging  disturbances in the jet which are likely to be  responsible for the high levels of $\gamma$-ray emission.

\end{itemize}

\acknowledgments
We thank the anonymous referee for careful reading and helpful comments. The Mets\"ahovi team acknowledges the support from the Academy of Finland (project numbers 212656, 210338, 121148, and others). We acknowledge the use of data and software facilities from the ASI Science Data Center (ASDC), managed by the Italian Space Agency (ASI).

\bibliography{ms}

\begin{deluxetable}{cccccccccccccccccccccccccccccccc} 
\rotate
\tablewidth{0pt}
\tabletypesize{\tiny}
\setlength{\tabcolsep}{0.03in} 
\tablecolumns{17}
\tablecaption{Luminosities \label{tab:tau}}
\tablehead{
\colhead{J2000\_Name} & \colhead{Alias}  & \colhead{RA} & \colhead{DEC}    & \colhead{$z$}   
                                 & \colhead{log L$_{30}$}       &  \colhead{log L$_{44}$} & \colhead{log L$_{70}$}   & \colhead{log L$_{100}$} 
                                 & \colhead{log L$_{143}$}     & \colhead{log L$_{217}$} & \colhead{log L$_{353}$}    & \colhead{log L$_{545}$} & \colhead{log L$_{857}$}&\colhead{$<log L_{\gamma}>_{sim}$}&\colhead{$< log L_{\gamma}>_{qua}$}&\colhead{$<log L_{\gamma}>_{ave}$}                                
 }
\startdata
     0010+1058&IIIZW2              &    2.629&   10.975& 0.089&    43.14&    43.30&    43.33&    43.45&    43.44&    43.45&    43.53&    43.99&$<$ 44.37&$<$ 45.34&$<$ 44.94&$<$ 44.42\\
   0017+8135&S50014+813          &    4.285&   81.586& 3.387&$<$ 45.59&$<$ 45.90&$<$ 46.04&$<$ 45.92&$<$ 45.84&$<$ 46.00&$<$ 47.09&$<$ 47.37&$<$ 47.82&$<$ 48.98&$<$ 48.75&$<$ 48.25\\
   0048+3157&Mkn348              &   12.196&   31.957& 0.015&$<$ 41.00&$<$ 41.35&$<$ 41.38&$<$ 41.22&$<$ 41.18&    41.40&$<$ 41.94&$<$ 42.44&$<$ 42.56&$<$ 43.46&$<$ 43.11&$<$ 41.91\\
   0120-2701&1Jy0118-272         &   20.132&  -27.023& 0.557&    44.25&$<$ 44.49&$<$ 44.60&    44.71&    44.66&    44.82&    45.06&$<$ 45.41&$<$ 45.63&$<$ 47.18&    46.86&    46.77\\
   0136+4751&S40133+47           &   24.244&   47.858& 0.859&    45.40&    45.51&    45.67&    45.80&    45.82&    45.93&    46.06&    46.19&$<$ 46.23&$<$ 47.63&    47.43&    47.51\\
   0204-1701&PKS0202-17          &   31.240&  -17.022& 1.740&    45.32&$<$ 45.48&    45.59&    45.65&    45.61&    45.48&$<$ 46.14&$<$ 46.64&$<$ 47.05&$<$ 48.30&$<$ 47.94&    47.80\\
   0210-5101&PKS0208-512         &   32.693&  -51.017& 1.003&    45.21&    45.42&    45.56&    45.59&    45.62&    45.65&    45.65&$<$ 45.85&$<$ 46.07&$<$ 47.65&    47.30&    47.51\\
   0214+5144&GB6J0214+5145       &   33.575&   51.748& 0.049&$<$ 41.93&$<$ 42.29&$<$ 42.48&$<$ 42.27&$<$ 42.11&$<$ 42.34&$<$ 42.98&$<$ 43.47&$<$ 43.89&\nodata&$<$ 44.17&$<$ 43.30\\
   0217+0144&PKS0215+015         &   34.454&    1.747& 1.715&    45.65&    45.78&    46.00&    46.13&    46.15&    46.30&    46.41&    46.48&$<$ 46.69&$<$ 48.63&    48.38&    48.17\\
   0217+7349&1Jy0212+735         &   34.378&   73.826& 2.367&    46.06&    46.08&    46.19&    46.24&    46.19&    46.27&$<$ 46.83&$<$ 47.19&$<$ 47.56&\nodata&$<$ 48.40&    48.23\\
   0221+3556&1Jy0218+357         &   35.273&   35.937& 0.944&    44.90&    44.99&$<$ 45.04&    45.22&    45.21&    45.28&    45.57&$<$ 46.35&$<$ 46.23&$<$ 48.02&    47.37&    47.60\\
   0237+2848&4C28.07             &   39.468&   28.802& 1.213&    45.38&    45.55&    45.62&    45.73&    45.72&    45.69&    45.82&$<$ 46.19&$<$ 46.75&$<$ 48.40&    48.02&    47.95\\
   0238+1636&PKS0235+164         &   39.662&   16.616& 0.940&    44.90&    45.09&    45.38&    45.53&    45.54&    45.62&    45.71&$<$ 45.87&$<$ 46.33&$<$ 48.06&$<$ 47.54&    47.99\\
   0319+4130&NGC1275             &   49.950&   41.512& 0.018&    42.54&    42.61&    42.68&    42.80&    42.83&    42.87&    42.90&    43.01&    43.10&\nodata&    44.38&    44.19\\
   0336+3218&NRAO140             &   54.125&   32.308& 1.259&    45.41&    45.59&    45.71&    45.78&    45.85&    45.82&$<$ 46.32&$<$ 46.60&$<$ 47.05&$<$ 48.04&$<$ 48.09&    47.64\\
   0423-0120&PKS0420-01          &   65.816&   -1.342& 0.916&    45.70&    45.87&    45.97&    46.12&    46.21&    46.25&    46.30&    46.44&    46.54&$<$ 48.03&    47.72&    47.63\\
   0428-3756&PKS0426-380         &   67.168&  -37.939& 1.030&    45.39&    45.52&    45.59&    45.48&    45.58&    45.60&    45.75&$<$ 45.99&$<$ 46.19&$<$ 47.70&    47.77&    48.28\\
   0433+0521&3C120               &   68.296&    5.354& 0.033&    42.26&    42.33&    42.35&    42.59&    42.66&    42.63&$<$ 42.66&$<$ 43.14&$<$ 43.60&$<$ 44.29&$<$ 43.93&$<$ 43.62\\
   0457-2324&PKS0454-234         &   74.263&  -23.414& 1.003&    45.22&    45.36&    45.49&    45.58&    45.65&    45.66&    45.76&    46.02&$<$ 46.21&    48.12&    47.95&    48.18\\
   0522-3627&PKS0521-36          &   80.741&  -36.458& 0.055&    43.03&    43.13&    43.33&    43.50&    43.55&    43.68&    43.71&    43.93&    44.01&$<$ 44.47&    44.39&    44.68\\
   0530+1331&PKS0528+134         &   82.735&   13.532& 2.070&    45.76&    46.01&    45.98&    46.08&    46.12&    46.14&$<$ 46.28&$<$ 46.77&$<$ 47.19&$<$ 48.93&$<$ 48.55&    48.47\\
   0538-4405&PKS0537-441         &   84.710&  -44.086& 0.892&    45.66&    45.80&    46.02&    46.16&    46.19&    46.33&    46.45&    46.58&    46.61&    48.45&    48.43&    48.21\\
   0539-2839&1Jy0537-286         &   84.976&  -28.665& 3.104&    45.87&    46.04&$<$ 46.02&    46.15&    46.05&    46.04&$<$ 46.43&$<$ 46.82&$<$ 47.49&$<$ 49.07&$<$ 48.99&    48.81\\
   0550-3216&PKS0548-322         &   87.669&  -32.272& 0.069&$<$ 42.35&$<$ 42.68&$<$ 42.78&$<$ 42.59&$<$ 42.46&$<$ 42.68&$<$ 43.25&$<$ 43.52&$<$ 43.90&$<$ 45.87&$<$ 45.23&    43.84\\
   0623-6436&IRAS-L06229-643     &   95.782&  -64.606& 0.129&    43.00&    43.37&    43.25&    43.25&$<$ 43.06&$<$ 43.19&$<$ 44.28&$<$ 44.56&$<$ 45.01&$<$ 44.93&$<$ 45.12&$<$ 44.84\\
   0738+1742&PKS0735+17          &  114.530&   17.705& 0.424&    44.08&$<$ 44.33&$<$ 44.44&$<$ 44.15&$<$ 44.13&$<$ 44.27&$<$ 45.36&$<$ 45.63&$<$ 46.09&$<$ 46.97&    46.54&    46.59\\
   0746+2549&B2.20743+25         &  116.608&   25.817& 2.979&$<$ 45.63&$<$ 45.97&$<$ 46.05&$<$ 45.79&$<$ 45.77&$<$ 45.91&$<$ 47.00&$<$ 47.28&$<$ 47.73&$<$ 49.49&$<$ 48.34&    48.56\\
   0818+4222&S40814+425          &  124.567&   42.379& 0.530&    44.59&    44.64&$<$ 44.64&$<$ 44.35&$<$ 44.33&$<$ 44.47&$<$ 45.56&$<$ 45.84&$<$ 46.29&$<$ 47.25&    46.69&    46.94\\
   0824+5552&OJ535               &  126.197&   55.878& 1.417&$<$ 45.01&$<$ 45.39&$<$ 45.47&$<$ 45.25&    45.35&$<$ 45.41&$<$ 45.95&$<$ 46.40&$<$ 46.59&$<$ 48.06&$<$ 47.85&    47.57\\
   0841+7053&4C71.07             &  130.351&   70.895& 2.218&    46.01&    46.23&    46.32&    46.56&    46.55&    46.56&    46.60&    46.69&$<$ 46.85&$<$ 48.47&$<$ 48.60&    48.49\\
   0854+2006&PKS0851+202         &  133.703&   20.108& 0.306&    44.76&    44.94&    45.11&    44.91&    44.98&    45.04&    45.16&    45.35&    45.43&$<$ 46.20&    45.98&    46.16\\
   0920+4441&S40917+44           &  140.243&   44.698& 2.190&    45.95&    46.03&    46.19&    46.33&    46.40&    46.47&    46.57&    46.75&    46.92&    48.88&    48.69&    48.76\\
   0923-2135&PKS0921-213         &  140.912&  -21.596& 0.053&$<$ 41.99&$<$ 42.39&$<$ 42.41&$<$ 42.36&    42.42&    42.54&    42.96&$<$ 43.48&$<$ 43.66&\nodata&$<$ 44.25&$<$ 44.00\\
   0957+5522&4C55.17             &  149.409&   55.383& 0.896&    44.89&$<$ 44.99&$<$ 45.06&    45.03&    45.07&    45.09&$<$ 45.45&$<$ 46.08&$<$ 46.19&    47.75&    47.74&    47.66\\
   1015+4926&1H1013+498          &  153.767&   49.433& 0.212&$<$ 43.35&$<$ 43.69&$<$ 43.78&$<$ 43.50&$<$ 43.46&$<$ 43.59&$<$ 44.25&$<$ 44.71&$<$ 44.90&    45.90&    45.91&    45.99\\
   1058+0133&4C01.28             &  164.623&    1.566& 0.888&    45.58&    45.71&    45.82&    45.94&    45.98&    46.06&    46.08&    46.19&    46.24&$<$ 47.69&    47.39&    47.51\\
   1058+5628&1RXSJ105837.5+562816&  164.657&   56.470& 0.143&$<$ 42.96&$<$ 43.32&$<$ 43.40&$<$ 43.24&$<$ 43.11&$<$ 43.21&$<$ 43.90&$<$ 44.36&$<$ 44.54&    45.67&    45.56&    45.38\\
   1058-8003&PKS1057-79          &  164.680&  -80.065& 0.581&    44.73&    44.91&    45.04&    45.21&    45.18&    45.14&    45.23&$<$ 45.44&$<$ 46.37&$<$ 46.90&    46.68&    46.68\\
   1104+3812&Mkn421              &  166.114&   38.209& 0.030&$<$ 41.48&$<$ 41.95&$<$ 42.04&    41.89&$<$ 41.83&$<$ 41.99&$<$ 42.55&$<$ 43.04&$<$ 43.46&    44.95&    44.98&    44.92\\
   1127-1857&PKS1124-186         &  171.768&  -18.955& 1.048&    45.15&    45.25&    45.24&    45.53&    45.60&    45.70&    45.88&$<$ 46.01&$<$ 46.38&    47.82&    47.86&    47.64\\
   1130-1449&PKS1127-145         &  172.529&  -14.824& 1.184&    45.60&    45.74&    45.86&    45.96&    45.97&    45.98&    46.10&    46.25&    46.41&$<$ 47.91&$<$ 47.73&    47.68\\
   1131+3114&B21128+31           &  172.789&   31.235& 0.289&$<$ 43.63&$<$ 43.97&$<$ 44.06&$<$ 43.84&$<$ 43.74&$<$ 43.96&$<$ 44.56&$<$ 45.06&$<$ 45.47&\nodata&\nodata&$<$ 45.09\\
   1136+7009&S51133+704          &  174.110&   70.157& 0.045&$<$ 41.98&$<$ 42.31&$<$ 42.36&$<$ 42.19&$<$ 42.14&$<$ 42.31&$<$ 42.90&$<$ 43.40&$<$ 43.82&$<$ 44.87&    44.04&    43.79\\
   1147-3812&PKS1144-379         &  176.755&  -38.203& 1.048&    45.05&$<$ 45.15&    45.30&    45.74&    45.80&    45.91&    46.03&    46.13&$<$ 46.28&$<$ 47.18&$<$ 46.83&    47.13\\
   1153+4931&4C49.22             &  178.352&   49.519& 0.334&    43.93&    44.17&    44.26&    44.28&    44.30&    44.24&$<$ 44.69&$<$ 45.19&$<$ 45.61&$<$ 46.16&$<$ 45.83&    45.65\\
   1159+2914&4C29.45             &  179.883&   29.246& 0.729&    44.75&    44.94&    44.89&    45.33&    45.39&    45.47&    45.58&    45.80&$<$ 45.85&    47.57&    47.46&    47.41\\
   1217+3007&ON325               &  184.467&   30.117& 0.130&    42.80&$<$ 43.25&$<$ 43.29&$<$ 43.09&$<$ 43.02&    43.22&$<$ 43.65&$<$ 44.10&$<$ 44.35&$<$ 45.73&    45.40&    45.46\\
   1220+0203&PKS1217+02          &  185.049&    2.062& 0.241&$<$ 43.96&$<$ 43.80&$<$ 44.28&    43.91&    43.89&    43.86&$<$ 44.37&$<$ 44.83&$<$ 45.01&\nodata&$<$ 45.91&$<$ 45.24\\
   1221+2813&ON231               &  185.382&   28.233& 0.102&$<$ 42.55&$<$ 42.96&$<$ 43.04&    42.90&    43.00&    43.10&$<$ 43.42&$<$ 43.84&$<$ 44.54&$<$ 45.59&    45.09&    45.19\\
   1222+0413&PKS1219+04          &  185.594&    4.221& 0.967&    45.04&$<$ 45.30&$<$ 45.25&    45.36&    45.35&    45.49&$<$ 45.66&$<$ 46.08&$<$ 46.26&    47.95&    47.60&    47.31\\
   1229+0203&3C273               &  187.278&    2.052& 0.158&    44.63&    44.85&    44.92&    45.09&    45.12&    45.08&    45.14&    45.25&    45.29&    46.58&    46.37&    46.35\\
   1246-2547&PKS1244-255         &  191.695&  -25.797& 0.635&    44.50&    44.75&    45.00&    45.12&    45.21&    45.27&    45.49&$<$ 45.77&$<$ 45.95&$<$ 47.68&    47.08&    47.31\\
   1256-0547&3C279               &  194.046&   -5.789& 0.536&    45.50&    45.66&    45.78&    45.94&    45.98&    46.06&    46.15&    46.28&    46.24&    47.60&    47.52&    47.64\\
   1305-1033&1Jy1302-102         &  196.387&  -10.555& 0.286&$<$ 43.54&$<$ 43.95&$<$ 44.06&$<$ 43.88&    44.01&    44.12&$<$ 44.46&$<$ 44.90&$<$ 45.46&\nodata&$<$ 45.97&$<$ 44.91\\
   1310+3220&1Jy1308+326         &  197.619&   32.345& 0.997&    45.46&    45.55&    45.55&    45.68&    45.70&    45.72&    45.83&    45.97&$<$ 46.15&    47.00&    47.28&    47.56\\
   1350+0940&GB6B1347+0955       &  207.592&    9.669& 0.133&$<$ 42.93&$<$ 43.27&$<$ 43.36&$<$ 43.17&$<$ 43.04&$<$ 43.27&$<$ 43.84&$<$ 44.29&$<$ 44.48&$<$ 45.01&$<$ 45.14&$<$ 44.04\\
   1419+0628&3C298.0             &  214.784&    6.476& 1.437&$<$ 44.95&$<$ 45.36&$<$ 45.48&$<$ 45.30&$<$ 45.16&$<$ 45.39&$<$ 45.96&$<$ 46.41&$<$ 46.60&$<$ 48.02&$<$ 47.75&$<$ 47.23\\
   1423+5055&BZQJ1423+5055       &  215.809&   50.927& 0.286&$<$ 43.47&$<$ 43.80&$<$ 43.90&    43.86&$<$ 43.68&$<$ 43.95&$<$ 44.53&$<$ 44.98&$<$ 45.17&$<$ 45.52&$<$ 44.98&$<$ 45.16\\
   1456+5048&1RXSJ145603.4+504825&  224.015&   50.807& 0.479&$<$ 44.09&$<$ 44.42&$<$ 44.26&$<$ 44.31&$<$ 44.20&$<$ 44.42&$<$ 44.61&$<$ 45.07&$<$ 45.36&$<$ 46.22&\nodata&$<$ 45.80\\
   1504+1029&PKS1502+106         &  226.104&   10.494& 1.839&    45.47&$<$ 45.59&    45.71&    45.74&    45.69&    45.74&$<$ 45.97&$<$ 46.61&$<$ 46.80&    48.60&    48.61&    49.16\\
   1507+0415&BZQJ1507+0415       &  226.999&    4.253& 1.701&$<$ 45.19&$<$ 45.53&$<$ 45.62&$<$ 45.36&$<$ 45.34&$<$ 45.48&$<$ 46.57&$<$ 46.85&$<$ 47.30&\nodata&$<$ 47.71&$<$ 46.46\\
   1510-0543&4C-05.64            &  227.723&   -5.719& 1.191&    45.03&$<$ 45.23&    45.44&$<$ 45.11&$<$ 45.00&$<$ 45.23&$<$ 45.83&$<$ 46.32&$<$ 46.74&$<$ 48.31&    47.66&    47.53\\
   1517-2422&APLib               &  229.424&  -24.372& 0.048&    42.52&    42.69&    42.83&    42.97&    43.05&    43.13&    43.33&    43.52&    43.78&$<$ 45.01&    44.66&    44.53\\
   1603+1554&WE1601+16W3         &  240.908&   15.901& 0.110&$<$ 42.77&$<$ 43.10&$<$ 43.19&$<$ 42.93&$<$ 42.91&$<$ 43.05&$<$ 44.14&$<$ 44.42&$<$ 44.87&$<$ 44.37&$<$ 44.25&$<$ 44.05\\
   1625-2527&OS-237.8            &  246.445&  -25.461& 0.786&    45.12&$<$ 45.11&    45.22&    45.37&    45.32&    45.44&$<$ 45.94&$<$ 46.75&$<$ 47.39&$<$ 47.95&    47.64&    47.61\\
   1635+3808&4C38.41             &  248.814&   38.134& 1.814&    45.93&$<$ 45.58&    46.23&$<$ 45.46&$<$ 45.35&$<$ 45.58&$<$ 46.17&$<$ 46.67&$<$ 47.09&    49.07&    48.99&    48.74\\
   1640+3946&NRAO512             &  250.123&   39.779& 1.660&$<$ 45.17&$<$ 45.51&$<$ 45.60&$<$ 45.69&    45.72&$<$ 45.92&$<$ 45.97&$<$ 46.53&$<$ 46.72&    48.50&    48.61&    48.43\\
   1642+3948&3C345               &  250.745&   39.810& 0.593&    45.39&$<$ 44.62&    45.60&    45.72&$<$ 44.39&$<$ 44.62&$<$ 45.21&$<$ 45.71&$<$ 46.13&    47.03&    47.14&    47.23\\
   1653+3945&Mkn501              &  253.468&   39.760& 0.033&$<$ 41.99&$<$ 42.03&$<$ 42.08&    42.13&    42.20&    42.20&    42.47&$<$ 42.75&$<$ 43.05&    44.62&    44.46&    44.46\\
   1719+4858&ARP102B             &  259.810&   48.980& 0.024&$<$ 41.42&$<$ 41.76&$<$ 41.85&    41.61&    41.46&$<$ 41.58&$<$ 42.04&$<$ 42.78&$<$ 42.78&\nodata&$<$ 43.51&$<$ 43.01\\
   1743+1935&1ES1741+196         &  265.991&   19.586& 0.084&$<$ 42.53&$<$ 42.86&$<$ 42.95&$<$ 42.74&$<$ 42.63&$<$ 42.86&$<$ 43.45&$<$ 43.95&$<$ 44.37&$<$ 44.63&$<$ 45.17&    44.27\\
   1751+0939&OT081               &  267.887&    9.650& 0.322&    44.69&    44.81&    44.95&    44.97&    45.00&    45.13&    45.18&    45.37&$<$ 45.49&$<$ 46.71&$<$ 46.20&    46.31\\
   1800+7828&S51803+784          &  270.190&   78.468& 0.680&    44.92&    45.05&    45.10&    45.33&    45.39&    45.50&    45.64&    45.81&    45.78&    47.16&    47.02&    47.01\\
   1833-2103&PKSB1830-210        &  278.416&  -21.061& 2.507&    46.26&    46.35&    46.40&    46.56&    46.54&    46.58&    46.61&$<$ 47.12&$<$ 47.60&    49.79&    49.79&    49.48\\
   1840-7709&PKS1833-77          &  280.160&  -77.158& 0.018&$<$ 41.17&$<$ 41.50&$<$ 41.60&$<$ 41.39&$<$ 41.28&$<$ 41.50&$<$ 42.10&$<$ 42.60&$<$ 43.01&$<$ 43.58&$<$ 42.78&$<$ 42.83\\
   1911-2006&2E1908.2-2011       &  287.790&  -20.115& 1.119&    45.45&    45.64&    45.73&    45.82&    45.91&    45.98&    46.00&    46.26&    46.54&$<$ 48.17&    47.65&    47.84\\
   1923-2104&PMNJ1923-2104       &  290.884&  -21.076& 0.874&    45.13&    45.38&    45.53&$<$ 44.79&    45.63&    45.72&    45.83&    45.99&$<$ 46.34&    47.74&    47.39&    47.58\\
   1924-2914&OV-236              &  291.213&  -29.242& 0.352&    45.18&    45.27&    45.39&$<$ 43.98&    45.41&    45.47&    45.55&    45.72&    45.64&$<$ 46.63&    46.06&    46.33\\
   1959+6508&1ES1959+650         &  299.999&   65.148& 0.047&$<$ 42.01&$<$ 42.35&$<$ 42.44&$<$ 42.23&$<$ 41.92&$<$ 42.06&$<$ 42.94&$<$ 43.34&$<$ 43.96&\nodata&\nodata&    44.48\\
   2009-4849&1Jy2005-489         &  302.355&  -48.831& 0.071&$<$ 42.35&$<$ 42.75&$<$ 42.88&    42.93&    42.87&    42.95&    43.22&$<$ 43.80&$<$ 43.86&$<$ 44.48&    44.64&    44.72\\
   2056-4714&PKS2052-47          &  314.068&  -47.246& 1.491&    45.57&    45.76&    45.83&$<$ 45.25&$<$ 45.24&$<$ 45.37&$<$ 46.46&$<$ 46.74&$<$ 47.19&$<$ 48.70&    48.16&    48.35\\
   2129-1538&1Jy2126-158         &  322.300&  -15.645& 3.268&$<$ 45.63&$<$ 46.02&$<$ 46.12&$<$ 45.89&$<$ 45.84&$<$ 46.02&$<$ 46.59&$<$ 47.05&$<$ 47.23&\nodata&$<$ 48.69&$<$ 48.49\\
   2143+1743&S32141+17           &  325.898&   17.730& 0.213&    43.62&$<$ 43.62&$<$ 43.76&    43.63&    43.78&    43.78&$<$ 44.29&$<$ 44.78&$<$ 45.20&$<$ 46.30&    46.11&    46.08\\
   2147+0929&1Jy2144+092         &  326.792&    9.496& 1.113&    44.99&    45.22&    45.25&    45.45&    45.49&    45.61&    45.94&$<$ 46.14&$<$ 46.68&    48.10&    47.95&    47.77\\
   2148+0657&4C06.69             &  327.022&    6.961& 0.999&    45.62&    45.75&    45.77&    45.87&    45.91&    45.91&    46.03&    46.04&$<$ 46.36&$<$ 47.96&$<$ 47.51&    46.81\\
   2151-3027&PKS2149-307         &  327.981&  -30.465& 2.345&    45.51&    45.77&$<$ 45.87&    45.94&    45.90&    46.01&    46.20&$<$ 46.57&$<$ 46.99&$<$ 48.99&$<$ 48.61&    48.43\\
   2202+4216&BLLac               &  330.680&   42.278& 0.069&    43.10&    43.24&    43.38&    43.52&    43.61&    43.65&    43.83&$<$ 44.09&$<$ 44.19&    45.07&    45.21&    45.13\\
   2203+3145&4C31.63             &  330.812&   31.761& 0.295&    44.26&    44.37&    44.60&    44.74&    44.78&    44.80&    44.88&    45.02&$<$ 45.49&$<$ 46.17&$<$ 45.93&$<$ 45.44\\
   2207-5346&PKS2204-54          &  331.932&  -53.776& 1.215&    45.03&    45.24&$<$ 45.34&    45.37&    45.41&    45.54&$<$ 45.63&$<$ 46.03&$<$ 46.46&$<$ 47.87&$<$ 47.59&    47.27\\
   2209-4710&NGC7213             &  332.317&  -47.167& 0.006&$<$ 40.22&$<$ 40.53&$<$ 40.64&$<$ 40.38&$<$ 40.36&$<$ 40.50&$<$ 41.59&$<$ 41.87&$<$ 42.32&$<$ 42.45&\nodata&    41.77\\
   2229-0832&PKS2227-08          &  337.417&   -8.548& 1.560&    45.50&    45.69&    45.95&    45.97&    46.06&    46.11&    46.23&$<$ 46.32&$<$ 46.58&$<$ 48.19&    48.16&    48.35\\
   2230-3942&PKS2227-399         &  337.668&  -39.714& 0.318&    43.72&$<$ 44.02&    44.22&    44.08&    44.01&$<$ 44.03&    44.43&$<$ 45.08&$<$ 45.27&$<$ 46.06&$<$ 45.18&$<$ 44.92\\
   2232+1143&4C11.69             &  338.152&   11.731& 1.037&    45.59&    45.68&    45.74&    45.78&    45.80&    45.83&    45.86&$<$ 45.99&$<$ 46.30&$<$ 47.93&$<$ 47.42&    47.65\\
   2253+1608&3C454.3             &  343.490&   16.148& 0.859&    46.00&    46.27&    46.54&    46.72&    46.85&    46.97&    47.12&    47.31&    47.36&    49.54&    49.23&    48.83\\
   2303-1842&PKS2300-18          &  345.762&  -18.690& 0.129&    42.93&$<$ 43.25&$<$ 43.43&$<$ 43.15&$<$ 43.04&$<$ 43.17&$<$ 43.66&$<$ 44.02&$<$ 44.29&$<$ 45.66&$<$ 44.56&$<$ 43.90\\
   2327+0940&PKS2325+093         &  351.890&    9.669& 1.843&    45.58&    45.66&    45.65&    45.74&    45.87&    45.88&$<$ 46.16&$<$ 46.62&$<$ 46.80&$<$ 48.56&$<$ 48.29&    48.31\\
   2333-2343&PKS2331-240         &  353.480&  -23.728& 0.048&    42.26&    42.38&    42.52&    42.69&    42.86&    42.99&    42.95&    43.22&$<$ 43.43&$<$ 44.44&$<$ 44.26&$<$ 43.60\\
   2347+5142&1ES2344+514         &  356.770&   51.705& 0.044&$<$ 41.85&$<$ 42.26&$<$ 42.38&$<$ 42.16&$<$ 42.08&$<$ 42.31&$<$ 42.88&$<$ 43.38&$<$ 43.80&$<$ 44.85&$<$ 44.74&    43.99\\
\enddata

\tablecomments{ All luminosities  are in units of erg s$^{-1}$ and have been computed by using the following cosmology H$_{0}$=71 km s$^{-1}$ Mpc$^{-1}$, $\Omega_{m}=0.27$, and $\Omega_{\Lambda}=0.73$ } \label{tab:sample}

\end{deluxetable}

\begin{deluxetable}{cccccccccccccccccccccccccccccccc} 
\tablewidth{0pt}
\tabletypesize{\tiny} 
\tablecolumns{5}
\tablecaption{Results of the correlation analysis between the mm/sub-mm and average $\gamma$-rays luminosities \label{tab:tau}}
\tablehead{
\colhead{} &    &  & \multicolumn{3}{c}{$ <L_{\gamma}> _{ave}$} &   \colhead{}     \\ 
\cline{3-7}  \\
 & \colhead{$n$/ul$_{Planck}$/ul$_{Fermi}$}       & \colhead{$\tau_{ll,z}$}     & \colhead{P$_{\tau}$}    & \colhead{P$_{surrogate}$}   & \colhead{$x$}   \\
  & \colhead{(1)}       & \colhead{(2)}     & \colhead{(3)}    & \colhead{(4)}   & \colhead{(5)}   
 }
\startdata

\sidehead{\textbf{L$_{30~GHz}$}}
ALL&
98/31/23   &   0.38  &  $ <10^{-6} $& $< 10^{-4}  $  & 1.26\\
FSRQ&
51/10/10   &  0.40&    5.$\times 10^{-5}$&    $ < 10^{-4} $&1.45\\
BLLac&
27/14/3     &  0.34&    9$\times 10^{-5}$&    0.1 &1.09  \\

\sidehead{\textbf{L$_{44~GHz}$}}
ALL&
98/46/23&  0.29&    2$\times 10^{-7}$&    $ <10^{-4}$ &1.30  \\
FSRQ&
51/18/10&  0.33&    5$\times 10^{-4}$&    1$\times 10^{-3}$ & 1.46 \\
BLLac&
27/18/3&  0.23&    7$\times 10^{-4}$&    0.1 &  1.18\\

\sidehead{\textbf{L$_{70~GHz}$}}
ALL&
98/43/23& 0.30&    6$\times 10^{-8}$&   $ <10^{-4}$  &1.29  \\
FSRQ&
51/16/10& 0.35&    5$\times 10^{-5}$&    $ <10^{-4}$  & 1.46 \\
BLLac&
27/18/3& 0.24&    4$\times 10^{-3}$&     0.1  & 1.17 \\

\sidehead{\textbf{L$_{100~GHz}$}}
ALL&
98/33/23& 0.31&    2$\times 10^{-6}$&    $ <10^{-4}$ &1.21  \\
FSRQ&
51/13/10& 0.33&    7$\times 10^{-4}$&     2$\times 10^{-3}$ &1.30  \\
BLLac&
27/13/3& 0.25&    7$\times 10^{-3}$&     0.2 & 1.06 \\

\sidehead{\textbf{L$_{143~GHz}$}}
ALL&
98/31/23& 0.34&    1$\times 10^{-6}$&    $ <10^{-4}$ & 1.19 \\
FSRQ&
51/11/10& 0.37&    2$\times 10^{-4}$&    $ <10^{-4}$ &1.30  \\
BLLac&
27/14/3& 0.28&    5$\times 10^{-3}$&     0.1  & 1.03 \\

\sidehead{\textbf{L$_{217~GHz}$}}
ALL&
98/33/23& 0.32&    4$\times 10^{-7}$&    $ <10^{-4}$ &1.21  \\
FSRQ&
51/13/10& 0.34&    3$\times 10^{-4}$&    1$ \times 10^{-3}$ &1.31  \\
BLLac&
27/12/3& 0.28&    3$\times 10^{-3}$&    0.1  & 1.04  \\

\sidehead{\textbf{L$_{353~GHz}$}}
ALL&
98/51/23& 0.21&    1$\times 10^{-5}$&     $ <10^{-4}$ & 1.31 \\
FSRQ&
51/25/10& 0.24&    2$\times 10^{-3}$&    2$\times 10^{-3}$ & 1.41 \\
BLLac&
27/15/3& 0.25&    1$\times 10^{-3}$&     0.1  &  1.13\\

\sidehead{\textbf{L$_{545~GHz}$}}
ALL&
98/70/23& 0.13&    8$\times 10^{-4}$&    $ <10^{-4}$ &  1.36\\
FSRQ&
51/33/10& 0.17&    0.01&    6$\times 10^{-3}$ &  1.5\\
BLLac&
27/22/3& 0.11&   0.05&     0.2  &  1.23\\

\sidehead{\textbf{L$_{857~GHz}$}}
ALL&
98/83/23& 0.10&    3$\times 10^{-3}$&    $ <10^{-4}$ & 1.32 \\
FSRQ&
51/42/10& 0.11&    5$\times 10^{-2}$&    1$\times 10^{-3}$ & 1.36 \\
BLLac&
27/23/3& 0.08&    8$\times 10^{-2}$&     0.1  & 1.24 \\

\enddata

\tablecomments{Column (1): (number of sources / number of Planck upper limits / number of Fermi upper limits ).  Column (2)  $\tau_{ll,z}$: partial correlation coefficient. Column (3) $P_{\tau}$:   probability that non-correlation exists between luminosities after the common effect of  redshift  has been removed . Colum (4)  $P_{surrogate}$: probability that the correlation arises by chance (estimated by surrogate methods). Column (5):  slope  fitted to the luminosity-luminosity relation. }\label{tab:correlation}

\end{deluxetable}

\begin{deluxetable}{cccccccccccccccccccccccccccccccc} 
\tablewidth{0pt}
\tabletypesize{\tiny} 
\tablecolumns{9}
\tablecaption{Results of the correlation analysis between the mm/sub-mm and  $\gamma$-rays emission integrated over different periods of time \label{tab:tau}}
\tablehead{
\colhead{} &      \multicolumn{3}{c}{$ <L_{\gamma}> _{sim}$} &   & \multicolumn{3}{c}{$ <L_{\gamma}> _{qua}$} &   & \multicolumn{3}{c}{$ <L_{\gamma}> _{ave}$}   \\ 
\cline{2-4} \cline{6-8} \cline{10-12} \\
     & \colhead{$\tau_{ll,z}$}     & \colhead{P$_{\tau}$}    & \colhead{P$_{surrogate}$}&  & \colhead{$\tau_{ll,z}$}     & \colhead{P$_{\tau}$}    & \colhead{P$_{surrogate}$}   &  & \colhead{$\tau_{ll,z}$}     & \colhead{P$_{\tau}$}    & \colhead{P$_{surrogate}$}   
 }    
\startdata

L$_{30}$     & 0.45& 0.03&    3$\times 10^{-3}$ &  & 0.48& 0.02&    2$\times 10^{-3}$ & & 0.50& 0.01&    $ <10^{-4}$  \\
L$_{44}$     & 0.34& 0.06&    \nodata & &0.35& 0.04&    6$\times 10^{-3}$&& 0.39& 0.03&    1$\times 10^{-3}$\\
L$_{70}$     & 0.46& 0.02&    3$\times 10^{-3}$ && 0.46& 0.01&    $< 10^{-4 }$&& 0.47& 0.01&    $<  10^{-4}$\\
L$_{100}$   & 0.36& 0.06&    \nodata  && 0.39& 0.03&   1$\times 10^{-2}$  && 0.39& 0.03&    1$\times 10^{-2}$\\
L$_{143}$   & 0.38& 0.06&    \nodata && 0.41& 0.04&    7$\times 10^{-3}$&& 0.46& 0.02&    $<  10^{-4}$\\
L$_{217}$   & 0.34& 0.08&    \nodata && 0.36& 0.06&    \nodata&& 0.39& 0.04&    5$\times 10^{-3 }$\\
L$_{353}$   & 0.32& 0.08&   \nodata && 0.34& 0.06&   \nodata&& 0.38& 0.04&    2$\times 10^{-3 }$\\
L$_{545}$   & 0.32& 0.04&     $ <10^{-4}$ && 0.33& 0.03&    $ <10^{-4}$&& 0.35& 0.03&    $ <10^{-4}$\\
L$_{857}$   & 0.29& 0.04&    $ <10^{-4}$ & & 0.29& 0.03&   1$\times 10^{-3}$  && 0.26& 0.05&   $ <10^{-4}$ \\

\enddata
\tablecomments{ $\tau_{ll,z}$: partial correlation coefficient.  $P_{\tau}$:   probability that non-correlation exists between luminosities after the common effect of  redshift  has been removed .   $P_{surrogate}$: probability that the correlation arises by chance (estimated by surrogate methods).}\label{table:detected}\label{tab:sim}

\end{deluxetable}

\begin{deluxetable}{cccccccccccccccccccccccccccccccc} 
\tablewidth{0pt}
\tabletypesize{\tiny}
\tablecolumns{5}
\tablecaption{(Sub-)mm spectral fit parameters  \label{tab:fits}}
\tablehead{
\colhead{J2000 Name} & \colhead{$\alpha_{mm}$}  & \colhead{$\alpha_{submm}$} & \colhead{$\nu_{break}$}    & \colhead{Spectral class}\\ 
\colhead{} & \colhead{}  & \colhead{} & \colhead{[GHz]}    & \colhead{}   
   }
   \startdata
  0010+1058 &-0.61 & 0.58 &343.13 &  B \\
 0120-2701 &-0.09 &-0.03 &100.00 &  B \\
 0136+4751 &-0.24 &-0.52 &120.57 &  C \\
 0210-5101 &-0.06 &-0.86 & 83.40 &  C \\
 0217+0144 &-0.08 &-0.40 &142.82 &  C \\
 0217+7349 &-0.66 &-0.74 & 70.06 &  A \\
 0221+3556 &-0.52 & 0.24 &159.85 &  B \\
 0237+2848 &-0.35 &-0.88 &128.39 &  C \\
 0238+1636 & 0.21 &-0.60 &111.08 &  D \\
 0319+4130 &-0.50 &-0.76 &138.77 &  C \\
 0336+3218 &-0.18 &-0.56 & 91.92 &  C \\
 0423-0120 &-0.25 &-0.61 &197.82 &  C \\
 0428-3756 &-0.43 &-0.63 & 77.46 &  A \\
 0457-2324 &-0.37 &-0.40 &198.48 &  B \\
 0522-3627 &-0.09 &-0.42 &110.00 &  C \\
 0530+1331 & 0.47 &-0.75 & 59.04 &  D \\
 0538-4405 &-0.02 &-0.32 &130.00 &  C \\
 0539-2839 &-0.53 &-0.65 &142.97 &  A \\
 0841+7053 &-0.11 &-0.97 & 96.84 &  D \\
 0854+2006 &-0.05 &-0.52 & 81.45 &  B \\
 0920+4441 &-0.39 &-0.09 &303.81 &  B \\
 1058+0133 &-0.44 &-0.52 &217.81 &  A \\
 1058-8003 &-0.14 &-1.12 & 95.28 &  C \\
 1130-1449 &-0.44 &-0.33 &180.09 &  B \\
 1229+0203 &-0.13 &-0.96 &126.62 &  C \\
 1246-2547 & 0.34 &-0.46 & 83.23 &  D \\
 1256-0547 &-0.16 &-0.56 &107.54 &  C \\
 1310+3220 &-0.63 &-0.61 &195.47 &  A \\
 1504+1029 &-0.37 &-0.97 & 87.73 &  C \\
 1517-2422 &-0.24 & 0.04 &180.06 &  B \\
 1751+0939 &-0.46 &-0.57 &127.79 &  A \\
 1800+7828 &-0.30 &-0.35 &183.88 &  A \\
 1833-2103 &-0.42 &-0.61 &111.17 &  A \\
 1911-2006 &-0.40 & 0.39 &296.73 &  B \\
 1923-2104 & 0.11 &-0.48 & 78.37 &  D \\
 1924-2914 &-0.42 &-0.60 & 90.06 &  A \\
 2147+0929 &-0.17 & 0.00 &100.00 &  B \\
 2148+0657 &-0.59 &-0.68 &205.52 &  A \\
 2151-3027 &-0.03 &-0.57 & 99.98 &  C \\
 2202+4216 &-0.21 &-0.64 &113.32 &  C \\
 2203+3145 &-0.08 &-0.76 &109.95 &  C \\
 2207-5346 &-0.37 &-0.41 &124.33 &  B \\
 2229-0832 & 0.24 &-0.64 & 96.46 &  D \\
 2232+1143 &-0.56 &-0.92 & 99.93 &  C \\
 2253+1608 & 0.47 &-0.26 & 97.99 &  D \\
 2327+0940 &-0.70 &-0.82 &129.50 &  A \\
 2333-2343 &-0.13 &-0.09 &336.43 &  B \\
  
\enddata
\tablecomments{The (sub-)mm spectra have been approximated with the broken power law model  of equation (5). }
\end{deluxetable}

\end{document}